\documentclass{iopart}

\usepackage{iopams}
\usepackage{graphicx}
\usepackage{cite}

\usepackage{perpage}
\MakePerPage{footnote}

\usepackage[colorlinks]{hyperref}

\begin{document}
\title[]{Conditioned random walks and interaction-driven condensation}

\author{Juraj Szavits-Nossan$^1$, Martin R Evans$^1$ and Satya N Majumdar$^2$}

\address{$^1$ SUPA, School of Physics and Astronomy, University of Edinburgh, Peter Guthrie Tait Road, Edinburgh EH9 3FD, United Kingdom}
\address{$^2$ Laboratoire de Physique Th\'{e}orique et Mod\`{e}les Statistiques, UMR 8626, Universit\'{e} Paris Sud 11 and CNRS, B\^{a}timent 100, Orsay F-91405, France}

\ead{jszavits@staffmail.ed.ac.uk}
\ead{mevans@staffmail.ed.ac.uk}
\ead{majumdar@lptms.u-psud.fr}

\begin{abstract}
We consider a discrete-time continuous-space random walk under the constraints that the number of returns to the origin (local time) and the total area under the walk are fixed. We first compute the joint probability of an excursion having area $a$ and returning to the origin for the first time after time $\tau$. We then show how condensation occurs when the total area constraint is increased: an excursion containing a finite fraction of the area emerges. Finally we show how the phenomena generalises previously studied cases of condensation induced by several constraints and how it is related to interaction-driven condensation which allows us to explain the phenomenon in the framework of large deviation theory.
\end{abstract}

%\ams{} 

\submitto{Journal of Physics A: Mathematical and Theoretical}

\date{\today}

\maketitle

%===========================================================%
% 1. INTRODUCTION				 
%===========================================================%

\section{Introduction}
\label{intro}

Recent years have witnessed an increased interest in the mathematical theory of large deviations, putting the theory at the fore of research in statistical physics. Roughly speaking, large deviation theory is concerned with probabilities of sums of random variables which are far away from their expected (mean) value \cite{Ellis95}. In physics, these sums are typically extensive variables such as energy and mass \cite{Touchette09}, but can also be time-integrated observables such as currents \cite{BDGJL05,BD05,Derrida07}. In most cases with short-range correlations between the random variables, the probability density that a sum of random variables $S_N=\sum_{i=1}^{N} x_i$ takes a large value $S_N=aN$ (meaning that $a\neq \langle S_N\rangle/N$) is found to decay exponentially
\begin{equation}
P(S_N=aN)\asymp \rme^{-NI(a)},\qquad N\rightarrow\infty,
\label{large-deviation-principle}
\end{equation}
where $\asymp$ denotes asymptotic behaviour in the sense that
\begin{equation}
I(a)=-\lim_{N\rightarrow\infty}\textrm{ln}P(S_N=aN)/N
\end{equation}
is the rate function characterising the exponential decay. When (\ref{large-deviation-principle}) holds, $P(S_N=aN)$ is said to satisfy the \emph{large deviation principle}.

While exceptions to this principle in physics are rare (at least for short-range correlated variables), their occurrence indicates a strikingly different behaviour. A prominent example is that of condensation phenomenon, where one of the random variables takes a macroscopic fraction of the sum \cite{EH05,Zannetti14,Zannetti15}. The simplest realisation is found in sums of independent and identically distributed (iid) random variables whose common probability density $f(x_i)$ is heavy-tailed, i.e. has tails that decay slower than exponential. In that case it is well known that the probability density $P(S_N=aN)$ is dominated by a single contribution to the sum \cite{Linnik61,Nagaev69,regular-variation,UchaikinZolotarev,Tribelsky02,MEZ05},
\begin{equation}
P(S_N=aN)\approx Nf((a-a_c)N),\qquad N\rightarrow\infty
\label{anomalous-large-deviation}
\end{equation}
where $a_c=\mathbb{E}_f[x]$ is the mean with respect to $f(x_i)$ and the factor $N$ indicates that any of the random variables can take the role of the condensate.

In physics, condensation phenomena have been extensively studied in mass-transport models such as the zero-range process (ZRP) and several related models \cite{EH05,MKB98,GSS03,MEZ05,EMZ06,HMS09,AL11,EW14,AGL13,WCEB14}. These models comprise masses $m_i$ (discrete or continuous) that are exchanged locally between neighbouring sites, typically on a one-dimensional lattice. Despite non-trivial (and generally non-equilibrium) dynamics, under certain conditions the steady state in a class of mass-trasport models (such as in ZRP) takes a simple factorised form \cite{EH05},
\begin{equation}
P(m_1,\dots,m_L) = \frac{1}{Z_L(M)}\left[\prod_{i=1}^{L}f(m_i)\right]
\delta\left(\sum_{j=1}^L m_j-M\right)\;.
\label{fss}
\end{equation}
Here $f(m_i)$ is the single-site weight, the delta function ensures that the total mass $M_L \equiv \sum_{i=1}^L m_i$ is conserved and equal to $M$, and $Z_L(M)$ is the normalisation constant (partition function) given by
\begin{equation}
Z_L(M)= \int\left[\prod_{i=1}^{L}{\rm d}m_i\, f(m_i)\right]\, 
\delta\left(\sum_{j=1}^L m_j-M\right)\; ,
\label{PF1}
\end{equation}
where $\int $ denotes the integral (for continuous masses) or the sum (for discrete masses). The condensation take place provided $f(m_i)$ is heavy-tailed and the density $\rho=M/L$ is greater than $\rho_c$, which is given by \cite{MEZ05,EMZ06}
\begin{equation}
\rho_c = \frac{\int_{0}^{\infty} \rmd m\, m f(m)}{\int_{0}^{\infty}\rmd m f(m)}\;.
\end{equation}
The heavy-tailed condition required for condensation is that  $f(m)$ (modulo any exponentially decreasing dependence on $m$) should decay to zero faster than $m^{-2}$ but slower  than exponentially \cite{EMZ06}. Thus
\begin{equation}
f(m) \simeq  A m^{-\gamma}\quad \mbox{with}\quad \gamma > 2
\end{equation}
fulfils the criterion as does a stretched exponential distribution
\begin{equation}
f (m) \simeq A m^\beta \exp(-cm^\alpha)
\quad \mbox{with}\quad \alpha <1\;.
\end{equation}

Recently, we showed that iid random variables with a light-tailed distribution can also exhibit condensation if there is an additional constraint on their linear statistics, such as the variance $\sum_j m_{j}^{2}$ \cite{SNEM14a}. In that case $f(m_i)$ does not need to be heavy-tailed, since it acquires an additional factor through the second constraint that has a Weibull-like heavy tail \cite{SNEM14b}.

\begin{figure}[hbt]
	\centering\includegraphics[width=10cm]{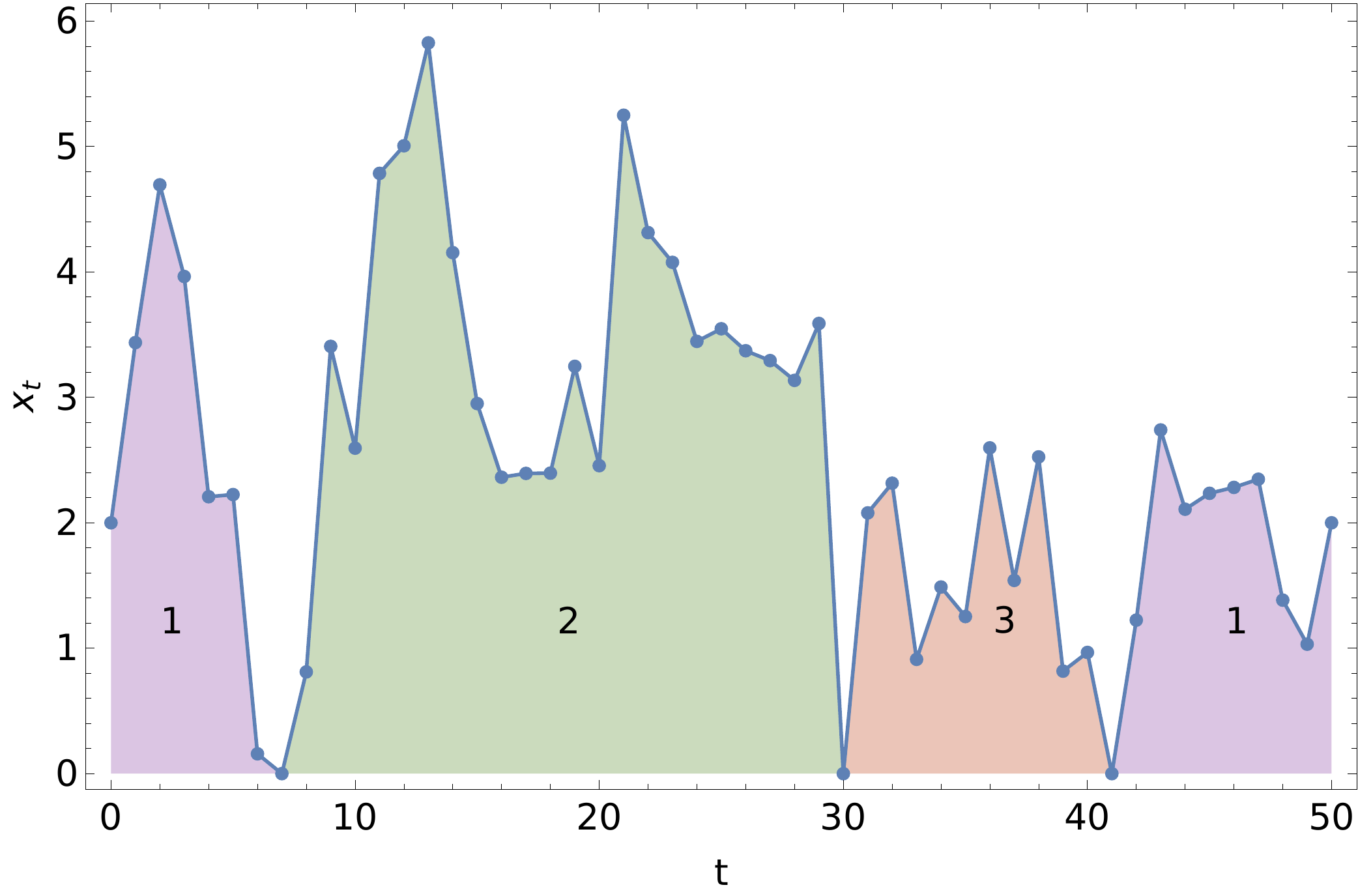}
	\caption{ A sample path for the reflected random walk with
		the double exponential (Laplace) jump distribution $K(\eta_t)=\textrm{exp}(-\vert\eta_t\vert)/2$. The path starts at $x_0=2$ at time $t=0$ and ends at $x_T=x_0=2$ at time $t=T=50$; it consists of three excursions denoted by numbers $1,2$ and $3$ that are represented by three different colours.}
	\label{fig1}
\end{figure}

In this paper we take a step further and study \emph{correlated} random variables conditioned on two constraints. As a prototype model belonging to this class, we study a discrete-time (continuous-space) reflected random walk on a line. A walker starts at the origin at time $t=0$, and undergoes independent jumps at each instant. If a jump takes it to the negative site, its position is reset to the origin (reflecting boundary condition) (see Figure \ref{fig1} for a typical sample path). In this model,  the relevant random variables are the positions $x_t$'s of the walker at different times. Even though the increments (jumps) at each instant are independent, the positions $x_t$'s at different times are strongly correlated. We are interested in paths of this process that realise fixed values of $A_T$ and $l_T$ defined by
\begin{equation}
A_T=\sum_{t=1}^{T}x_t,\qquad l_T=\sum_{t=1}^{T}\delta_{x_t,0},
\end{equation}
where $A_T$ is the area under the path and $l_T$ is the number of returns to the origin, which is also known as (discrete) local time. One of our main results is to show that these two constraints together can generate a single random walk excursion---a path between two successive returns to the origin---which takes a macroscopic fraction of the total area $A_T$. In this context, this single excursion is the analogue of a condensate. We will discuss when and how this form of condensation occurs.

This condensation phenomenon is seemingly different from the standard one studied for i.i.d. random variables in the literature \cite{EH05,MEZ05}, considering that the condensate spans  more than one random variable $x_t$. However, we show how to recast the process in terms of random walk excursions,  i.e., the paths between successive returns to the origin, which are mutually independent (due to the Markov nature of the process), thus connecting this condensation phenomenon to the standard one for iid random variables. Even though our conclusion holds for generic short-range jump distribution, in this paper we will present explicit results for the double exponential (or Laplace) jump distribution for which the transition line (between fluid and condensed phases) can be found exactly. Indeed, this particular jump distribution has played an important role already in computing exactly the probability distribution of several interesting variables, such as the area under an excursion of a discrete-time random walk \cite{SM06}. In our case also, it turns out that this special choice of the jump distribution makes analytical calculations possible.

Finally, we establish a direct connection between the conditioned random walk paths discussed above and the mass-transfer models in which the steady state assumes a pair-factorised form \cite{EHM06} given by
\begin{equation}
P(m_1,\dots,m_L)=\frac{1}{Z_{L}(M)}\prod_{i=1}^{L}g(m_{i},m_{i+1})\delta\left(\sum_{j=1}^{L}m_j-\rho L\right),
\label{pair-factorised-probability}
\end{equation}
where the pair function $g(m_{i},m_{i+1})$ consists of the interaction part $-J\vert m_{i+1}-m_i\vert$ and the single-site potential $U\delta_{m_i,0}$
\begin{equation}
g(m_i,m_{i+1})=\textrm{exp}\left[-J\vert m_{i+1}-m_{i}\vert+{1\over 2}U\delta_{m_i,0}+{1\over 2}U\delta_{m_{i+1},0}\right].
\label{pair-function}
\end{equation}

The most striking feature of the steady state (\ref{pair-factorised-probability}) is that it exhibits  spatially-extended (or interaction-driven) condensation for particle density $\rho$ above some critical density $\rho_c$, given in (\ref{rhoc}). However, a precise understanding of this novel type of condensation within the theory of large deviations is still lacking.

To fill this gap, we show that the phenomenon of interaction-driven condensation is related to the standard condensation, albeit one that is driven by two constraints rather than one. We arrive at this result by mapping the problem to a random walk that stays non-negative and using recent results on the equivalence of nonequilibrium path ensembles \cite{CT13,CT15}. Our results therefore unify these two seemingly disparate condensation phenomena and lead us to think that other, more complex condensation phenomena may have a similar origin \cite{Frey15,SNE15}.

The paper is organised as follows. In Section \ref{conditioned-random-walk} we study a discrete-time and continuous-space reflected random walk conditioned on the fixed total area $A_T$ and number of returns to the origin $l_T$. After recasting the process in terms of successive random walk excursions, in Section \ref{analysis-parition-function} we show how fixing $A_T$ causes a single excursion to take a finite fraction of the total area, thus signalling a condensation transition. In Section \ref{interaction-driven} we establish a direct connection between the conditioned random walk discussed above and the mass-transfer models with pair-factorised steady states that exhibit interaction-driven condensation. Finally we conclude in Section \ref{conclusion} with a summary and discussion. Some details are relegated to the Appendices.

%===============================================================================%
% 2. REFLECTED RANDOM WALK CONDITIONED ON AREA AND LOCAL TIME
%===============================================================================%

\section{Reflected random walk conditioned on area and local time}
\label{conditioned-random-walk}

We consider a discrete-time, continuous space random walk on a line. The walker starts at the origin at time $t=0$. In this model, when the position of the walker becomes negative, its position is instantaneously reset to the origin. The process, which we refer to as the {\it reflected random walk}, is described by the following recurrence equation
\begin{equation}
	x_{t}=\textrm{max}\{0,x_{t-1}+\eta_t\},
	\label{reflected-random-walk}
\end{equation}
where $x_t$ is position of the random walker at time step $t$ and $\eta_t$, $t=1,2,\dots$, are independent and identically distributed (iid) random variables with a common probability density $K(\eta_t)$. We note that the process (\ref{reflected-random-walk}) is well known in  queuing theory, where it is called the Lindley process \cite{Lindley52,Asmussen03}. Here we are interested in paths ${X}=(x_0,\dots,x_T)$ that start from $x_0=x$ at $t=0$ and return to the same position $x_T=x_0=x$ at time $T$. In the mathematical literature, these are known as stochastic bridges. For a reflected random walk, stochastic bridges coincide with stochastic excursions, which are defined as bridges that stay non-negative.

For a given path $X=\{x_0,\dots,x_T\}$, we are interested in the following two functionals of $X$: the total area $A_T[X]$
\begin{equation}
A_T[X]=\sum_{t=1}^{T}x_t,
\end{equation}
and the number of returns to the origin $l_T[X]$, also known as local time,
\begin{equation}
l_T[X]=\sum_{t=1}^{T}\delta_{x_t,0}.
\end{equation}
In addition, we are interested in all paths that achieve fixed values of $A_T=A$ and $l_T=N$, where both $A$ and $N$ are proportional to the total duration $T$ of the walk,
\begin{equation}
	A=(\sigma/\mu)T,\qquad N=(1/\mu)T,
	\label{A-N}
\end{equation}
and $\sigma$ and $\mu$ are positive constants. Generally we choose values of $\sigma$ and $\mu$ which correspond to large deviations of the unconstrained walk.

Our main goal in this paper is to show that conditioning paths on their total area and local time, both set proportional to the total duration $T$ of the walk, leads to the emergence of a single large excursion, which we refer to as the condensate, that takes a macroscopic fraction of the fixed total area $A$. This condensation phenomenon occurs when $\sigma$ (defined in (\ref{A-N})) is larger than the critical $\sigma_c$ (for a fixed $\mu$), which we calculate explicitly for the double exponential (also known as the Laplace) jump distribution $K(\eta)=\textrm{exp}(-\vert \eta\vert)/2$ (normalised to unity). We also show that conditioning on $l_T=N$ alone does not cause the transition---it is only when both $A_T$ and $l_T$ are fixed that the condensate emerges.

We start by writing down the path probability $P[X\vert A_T=A, l_T=N]$ of observing a path $X$ conditioned on the events $A_T=A$ and $l_T=N$, which reads
\begin{equation}
\fl\qquad P[X\vert A_T=A,l_T=N]=\frac{1}{Z_{N}(A,T)}\prod_{t=1}^{T}w(x_t\vert x_{t-1})\delta(l_T-N)\delta(A_T-A),
	\label{path-probability}
\end{equation}
where $Z_{N}(A,T)$ and $w(x_t\vert x_{t-1})$ are the normalisation constant and the transition probability, respectively,
\begin{equation}
\fl\qquad Z_{N}(A,T)=\int_{0}^{\infty}\rmd x_1\dots\int_{0}^{\infty}\rmd x_T\prod_{t=1}^{T}w(x_t\vert x_{t-1})\delta(l_T-N)\delta(A_T-A),
	\label{partition-function}
\end{equation}
\begin{equation}
w(x_t\vert x_{t-1})=\delta_{x_{t},0}\int_{-\infty}^{-x_{t-1}}K(\eta)\rmd\eta+K(x_t-x_{t-1}).
	\label{transition-probability}
\end{equation}

We note that the variable $x_t$ in (\ref{reflected-random-walk}) is neither a continuous nor a discrete random variable, but a mixture of both. Such mixed random variables are described by a generalised transition probability (\ref{transition-probability}) that involves the Dirac delta function. The form in (\ref{transition-probability}) takes into account the fact that if a jump at time $t-1$ takes the walker to the negative side, its position at $x_t$ is instantaneously reset to $0$.  

Instead of writing the delta function $\delta(l_T-N)$ in (\ref{partition-function}) and integrating over all paths, we can take $N$ returns explicitly into account, as follows. Let us denote with $t_i$, $i=1,\dots,N$, the time of the $i$-th return to the origin. Then the path between $t_{i-1}$ and $t_{i}$ is a random walk excursion and we let $\tau_i$ and $a_i$ denote its duration and area, 
respectively,
\begin{equation}
\tau_i=t_{i}-t_{i-1},\qquad a_i=\sum_{t=t_{i-1}}^{t_{i}}x_t.
\end{equation}
(Note that since $x_T=x_0$, the duration and area of the first excursion are given by $\tau_1=t_1+(T-t_{N})$ and $(x_{t_N}+\dots x_{T})+(x_1+\dots+x_{t_1})$, respectively; see Figure \ref{fig1} for more details.) The partition function $Z_{N}(A,T)$ can then be written in an alternative form
\begin{equation}
\fl\qquad Z_{N}(A,T)=
 \sum_{\{\tau_i =1\}}^{\infty}\prod_{i=1}^N \int_0^\infty {\rm d}a_i\,
f(a_i,\tau_i)\delta\left(\sum_{j=1}^{N}a_j-A\right)\delta\left(\sum_{k=1}^{N}\tau_{k}-T\right)
\label{partition-main}
\end{equation}
where $f(a_i,\tau_i)$ is the joint probability density for the duration $\tau_i$ and area $a_i$ of excursion $i$.

The expression (\ref{partition-main}) for the partition function $Z_N(A,T)$ is central to our paper. The advantage of introducing new random variables $a_i$ and $\tau_i$ is that the pairs $\{a_i,\tau_i\}$, for different $i=1,\dots,N$ are mutually independent (except for the global constraints on their sums), in contrast to the positions $x_t$, $t=1,\dots,T$, which depend on $x_{t-1}$. This formulation thus exploits explicitly the renewal nature of the process (see Figure \ref{fig1}).

Here we compute the joint proba\-bility density $f(a,\tau)$ for the discrete-time and continuous-space random walk with the double exponential jump probability density
\begin{equation}
	K(\eta_t)=\frac{1}{2}\rme^{- \vert \eta_t\vert}.
\label{de1}
\end{equation}

To the best of our knowledge, explicit calculations of $f(a,\tau)$ have been made so far only for the simple (Bernoulli) random walk (discrete time and space) by Tak\'{a}cs \cite{Takacs91} and for the Brownian motion (continuous time and space) by Kearney and Majumdar \cite{KM05}. In that context, our results present a novel calculation for the discrete-time and continuous-space random walk.

\subsection{Explicit calculation for the double exponential jump distribution}
\label{double-exponential}

To calculate $f(a,\tau)$, we need to study the probability of the following event. We consider a free random walker starting at the origin that makes successive jumps, stays positive till step $\tau-1$ and at time $\tau$ becomes negative for the first time. Thus $\tau$ is the first-passage time and is a random variable. Let $a=\sum_{t=0}^{\tau-1} x_t $ denote the `area' contained under the walk up to  this first-passage time. Then $f(a,\tau)$ denotes the joint probability density of the area $a$ till the first-passage time and the first-passage time $\tau$ itself. Problems where one is interested in computing distributions of functionals of a random walk till its first-passage time are generally referred to {\it first-passage functionals} \cite{M05}. To compute these distributions, it turns out to be advantageous to use a backward Fokker-Planck approach, suitably adapted \cite{M05,M10}. For this purpose, we will consider the starting position of the random walker $x_0=x\ge 0$ as a variable, calculate distributions of the functionals for a given $x\ge 0$, and eventually set $x=0$ to obtain our desired result. 

\begin{figure}[hbt]
\centering\includegraphics[width=10cm]{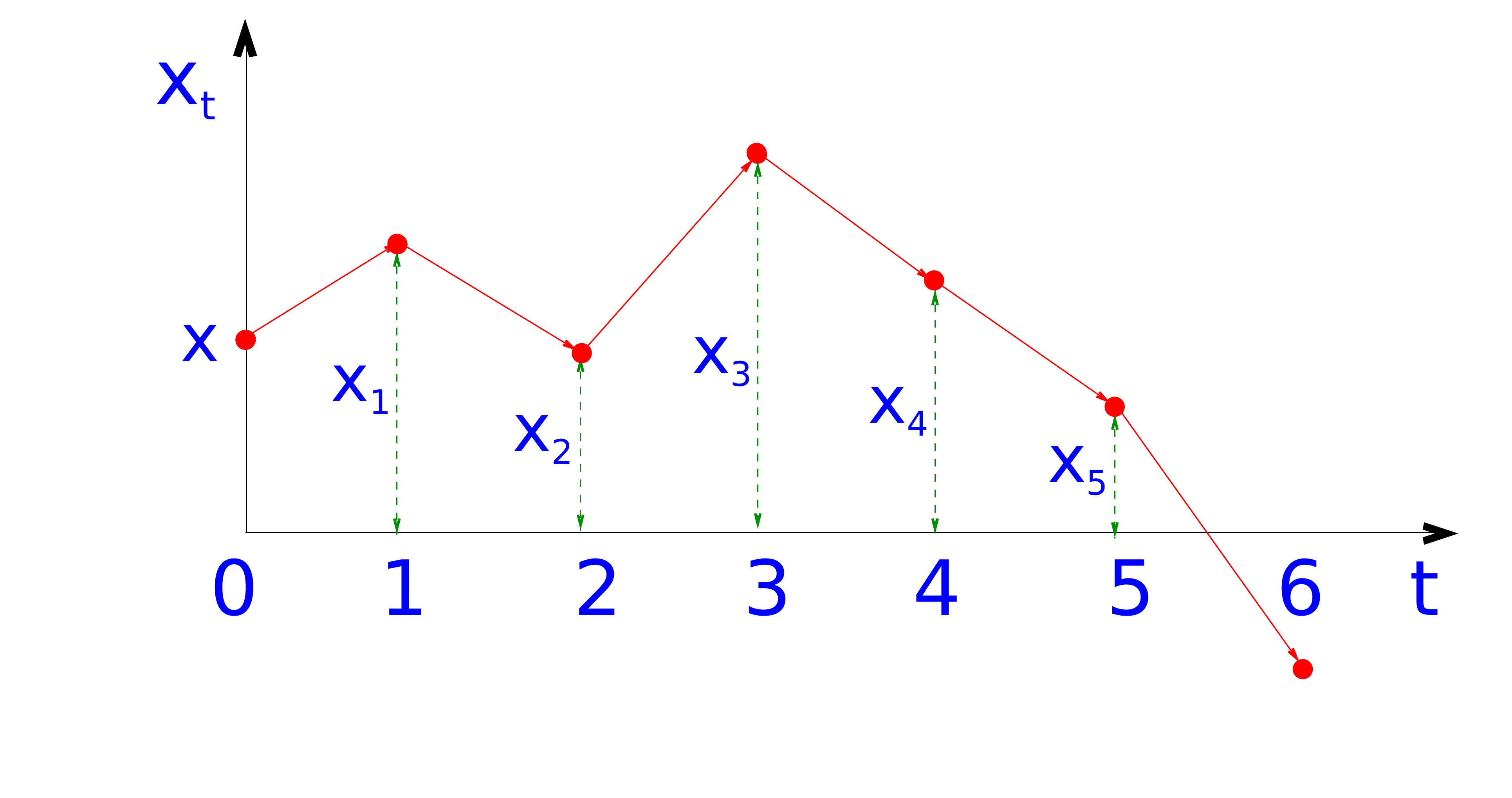}
\caption{A schematic sample path for the random walk that starts at the initial position $x_0=x\ge 0$ at time $t=0$ and its position evolves via successive independent jumps. It crosses $0$ to the negative side for the first time at $\tau=6$. The total area till the first-passage time $\tau=6$ is defined as
$a= \sum_{t=0}^{\tau-1=5} x_t$.} 
\label{fig2}
\end{figure}

Let $f(x,a,\tau)$ denote the joint probability density of $a$ and $\tau$, starting at $x_0=x$, where, as before,   $a= \sum_{t=0}^{\tau-1} x_t$. 
To find $f(x,a,\tau)$ we have to integrate over all positions 
$x_{t}$, $t=1,\dots,\tau$, where $x_t$ is strictly positive at $t=1,\dots,\tau-1$ and negative or zero at $t=\tau$ (see Figure 
\ref{fig2} for a schematic representation):
\begin{eqnarray}
\fl \qquad f(x,a,\tau)&=&\int_{0}^{\infty}\rmd x_1\dots \rmd x_{\tau-1}\;K(x_{1}-x)\cdot\dots\cdot K( x_{\tau-1}-x_{\tau-2})\nonumber\\
&&\times\int_{-\infty}^{0}\rmd x_\tau\; K(x_\tau-x_{\tau-1})\delta\left(\sum_{t=0}^{\tau-1}x_t-a\right)\nonumber\\
&=& \int_{0}^{\infty}\rmd x_1\;K(x_1-x)f(x_1,a-x,\tau-1),\quad \tau>1
\label{pgt1}
\end{eqnarray}
This integral equation has a simple interpretation: in the first step the walker jumps from the initial position $x$ to a new point at $x_1>0$ (the probability of this event is simply $K(x_1-x)\, dx_1$). Following this first step the process renews itself with $x_1$ being the new starting position. For the subsequent evolution starting at $x_1$, the area should be $a-x$ and the first-passage time should be $\tau-1$, explaining the factor $f(x_1, a-x, \tau-1)$ on the rhs. Finally, one needs to integrate over all possible $x_1\ge 0$. A separate case to consider is for $\tau=1$ for which $f(x,a,\tau)$ is given by
\begin{equation}
f(x,a,1)=\delta(x-a)\int_{-\infty}^{0}\rmd x_1 K(x_1-x).
\label{fpt1}
\end{equation}
This happens if at the first step the walker jumps to the negative 
side. Note that the recursion equations (\ref{pgt1}) and (\ref{fpt1}) hold for arbitrary jump distribution $K(\eta)$.
 
To make further progress, it is useful to consider the Laplace transform/moment-generating function $G(x,p,z)$ defined as
\begin{equation}
G(x,p,z)=\int_{0}^{\infty}\rmd a\;\rme^{-pa}\sum_{\tau=1}^{\infty}f(x,a,\tau)z^{\tau}.
\label{G}
\end{equation}
The equation for $G(x,p,z)$ is obtained by inserting (\ref{G}) into (\ref{pgt1}) and (\ref{fpt1}), yielding
\begin{eqnarray}
G(x,p,z)&=& z\rme^{-px}\int_{0}^{\infty}\rmd x_1\;K(x_1-x)G(x_1,p,z)\nonumber\\
&&+z\rme^{-px}\int_{-\infty}^{0}\rmd x_1 K(x_1-x).
\label{Geq}
\end{eqnarray}
These integral equations (where the limits of the integrals on the right hand side are over a semi-infinite line) are generally known as Weiner-Hopf equations and they are notoriously hard to solve explicitly for arbitrary jump distribution $K(\eta)$ (for a discussion see \cite{M10}). However, in the special case of the double exponential distribution (\ref{de1}), one can make progress, as we now show.

A key simplification for the double exponential case that allows us to solve the above integral equation is the following identity
\begin{equation}
\frac{\rmd^2}{\rmd x^2}\rme^{-\vert y-x\vert}=\rme^{-\vert y-x\vert}-2\delta(x-y).
\label{exp_id}
\end{equation}
By taking the second derivative of (\ref{Geq}) and making use of the identity (\ref{exp_id}) gives the following differential equation for $G(x,p,z)$
\begin{equation}
\frac{\rmd^2 G}{\rmd x^2}+2p\frac{\rmd G}{\rmd x}+(p^2-1+z\rme^{px})G=0.
\label{diff.1}
\end{equation}
In order to reduce this second order ordinary differential equation to a standard form, we first make a change of variable $u=2z^{1/2}\rme^{-px/2}/p$ and then define $W(u,p,z) = \rme^{px} G(x,p,z)$. It is then easy to see that $W(u)$ satisfies Bessel's differential equation
\begin{equation}
u^2\frac{\rmd^2W}{\rmd u^2}+u\frac{\rmd W}{\rmd u}+[u^2-(2/p)^2]W=0.
\label{diff.Bessel}
\end{equation}
Hence, the general solution to the original differential equation (\ref{diff.1}) can be written as
\begin{equation}
\fl \qquad W(x,p,z)=A(p,z)\, J_{2/p}\left(2z^{1/2}\rme^{-px/2}/p\right) + B(p,z)\,  N_{2/p}\left(2z^{1/2}\rme^{-px/2}/p\right),
\end{equation}
where $J_{\nu}$ and $N_{\nu}$ are Bessel functions of the first and second kind respectively, and $A$ and $B$ are arbitrary constants independent of $x$ (they are however functions of $p$ and $z$). Since $G$ must not diverge as $x\rightarrow\infty$, we can discard the second solution (set $B=0$) to obtain
\begin{equation}
G(x,p,z)=A(p,z)\, \rme^{-px}\, J_{2/p}\left(2z^{1/2}\rme^{-px/2}/p\right).
\label{sol.1}
\end{equation}
Determining the constant $A(p,z)$ is however far from trivial. To fix this, we substitute the solution (\ref{sol.1}) back into the original integral equation (\ref{Geq}), and after somewhat laborious algebra described in \ref{appendix_a}, we finally obtain $A(p,z)$ in the convenient closed form
\begin{equation}
A(p,z)=\frac{z^{1/2}}{J_{2/p{-}1}(2z^{1/2}/p)}.
\label{Apz.1}
\end{equation}

Equation (\ref{sol.1}) together with (\ref{Apz.1}) are the main result of this section. In particular, setting $x=0$ gives the desired Laplace transform/moment-generating function $g(p,z)$ for the joint probability density $f(a,\tau)=f(x=0,a,\tau)$
\begin{equation}
\fl\qquad g(p,z)\equiv G(x=0,p,z)\equiv \int_0^{\infty} \rmd a\, \rme^{-p\,a}\, \sum_{\tau=1}^{\infty} f(a,\tau)z^{\tau}
=\frac{z^{1/2} J_{2/p}(2z^{1/2}/p)}{ J_{2/p-1}(2z^{1/2}/p)}.
\label{transform}
\end{equation}
The Laplace transform/moment-generating function $g(p,z)$ in (\ref{transform}) is sufficient to show that the partition function $Z_N(A,T)$ exhibits a condensation transition, which we prove in the next Section.

For later usage in the next section, it is convenient to define at this point the inverse Laplace transform (with respect to $p$) of $g(p,z)$ in (\ref{transform})
\begin{equation} 
\widetilde f(a,z)= \sum_{\tau=1}^{\infty} f(a,\tau)\, z^\tau=
\int_{\gamma-i\infty}^{\gamma+ i \infty} \frac{{\rm d}p}{2\pi i}\, {\rm 
e}^{ap}\, g(p,z) 
\label{gtilde} 
\end{equation}
where the contour $\gamma$ is the usual Bromwich contour taken parallel to the imaginary axis and to the right of any singularities in the complex $p$ plane.

%===============================================================================%
% 3. ANALYSIS OF THE PARTITION FUNCTION AND THE CONDENSATION TRANSITION
%===============================================================================%
\section{Analysis of the partition function and the condensation transition}
\label{analysis-parition-function}

In this Section we analyse the partition function $Z_N(A,T)$ defined in (\ref{partition-main}) and the circumstances under which the condensation occurs. To this end, we study the Laplace transform/moment-generating function $\mathcal{Z}_N(p,z)$ defined as
\begin{equation}
\mathcal{Z}_N(p,z)=\sum_{T=0}^{\infty}z^{T}\int_{0}^{\infty}\rmd A\;\rme^{-pA}Z_N(A,T)=[g(p,z)]^N\;,
\label{Z_mgf_laplace}
\end{equation}
where $g(p,z)$ is given in (\ref{transform}). We first show that fixing \emph{either} the local time $l_T$ or the total area $A_T$, while allowing the other observable to fluctuate, does {\em not} lead to condensation.

\subsection{No condensation transition when either local time or total area is fixed}
\label{no-condensation-transition}

Fixing only the local time $l_T$ amounts to setting $p=0$ in (\ref{Z_mgf_laplace}) since
\begin{eqnarray}
\mathcal{Z}_N(p=0,z)&=&\sum_{T=0}^{\infty}z^{T}\int_{0}^{\infty}\rmd A\; Z_N(A,T)\nonumber\\
&=&\sum_{T=0}^{\infty}z^{T}Z_N(T)=[g(p=0,z)]^N,
\label{Z_mgf_laplace_p0}
\end{eqnarray}
where $Z_N(T)$ is given by
\begin{equation}
Z_N(T)=\sum_{\{\tau_i=1\}}^{\infty}\prod_{i=1}^N 
f_{\tau}(\tau_i)\, \delta\left(\sum_{j=1}^{N}\tau_{j}-T\right)
\label{ZNT.1}
\end{equation}
and $f_{\tau}(\tau)= \int_0^{\infty} f(a,\tau)\, \rmd a$ is the (marginal)  distribution of the first-passage time $\tau$. Also, $g(p=0,z)= \sum_{\tau=1}^{\infty} f_{\tau}(\tau)\, z^\tau$ is just the generating function of the first-passage time distribution.

Note that Eq. (\ref{ZNT.1}) is similar to the standard form of the partition function in (\ref{PF1}) with a single site weight $f_{\tau}(\tau_i)$. Hence, from the general criterion for condensation discussed in Section \ref{intro}, we expect condensation to happen if $f_{\tau}(\tau)$ decays for large $\tau$ slower than exponential (for example, algebraically with an exponent more than $2$). We show below that this is not the case.

To compute $f_{\tau}(\tau)$, we compute its generating function $g(p=0,z)$ by taking the limit $p\rightarrow 0$ of $g(p,z)$ defined in (\ref{transform}). Alternatively, we can compute $g(p=0,z)$ by going back to the equation for $G(x,p,z)$, which becomes much simpler when $p=0$ yielding
\begin{equation}
\frac{\rmd^2 G}{\rmd x}+(z-1)G=0.
\end{equation}
The general solution to this equation is 
\begin{equation}
G(x,0,z)=A_0(z)\rme^{-x\sqrt{1-z}}+B_0(z)\rme^{x\sqrt{1-z}},
\label{Gsol_p0}
\end{equation}
from  which we discard the second solution as  being nonphysical so that $B_0(z)=0$. The constant $A_0(z)$ can be then calculated by inserting (\ref{Gsol_p0}) into the integral equation for $G(x,p,z)$ in (\ref{Geq}) and following the same steps as before, which are detailed in \ref{appendix_a}. The final result for $G(x,0,z)$ is
\begin{equation}
G(x,0,z)=\left[1-\sqrt{1-z}\right]\rme^{-x\sqrt{1-z}},
\end{equation}
from which the moment-generating function of the first-passage time $\tau$ is obtained by letting $x=0$ yielding
\begin{equation}
g(p=0,z)=\sum_{\tau=1}^{\infty}f_{\tau}(\tau)\, z^{\tau}=1-\sqrt{1-z}.
\label{gz0.0}
\end{equation}
Expanding $g(0,z)$ around $z=0$ gives the following expression for the first-passage time distribution $f_{\tau}(\tau)$ and its tail for large $\tau$,
\begin{equation}
f_{\tau}(\tau)=\frac{1}{2^{2\tau-1}\tau}\left ({2\tau-2 \atop 
\tau-1}\right)\sim 
\frac{1}{2\sqrt{\pi}\tau^{3/2}},\qquad \tau\rightarrow\infty.
\label{f_marginal}
\end{equation}
The distribution $f_{\tau}(\tau)$ has a power-law tail with the exponent $3/2$,  which is less than the value $2$ required for condensation to happen. We thus conclude that fixing local time $l_T$ alone is not enough to induce condensation. 

Importantly, this conclusion extends to jump distributions other than the double exponential distribution studied here. It is due to the remarkable fact that the first-passage time distribution $f_{\tau}(\tau)$ in (\ref{f_marginal}) is universal for a discrete-time random walk with arbitrary continuous jump distribution, which is contained in the Sparre Andersen theorem \cite{SA54,Feller2,BMS13}.

In a similar fashion, we consider the case in which only the total area $A_T$ is fixed. This amounts to setting $z=1$ in (\ref{Z_mgf_laplace}) since
\begin{eqnarray}
\mathcal{Z}_N(p,z=1)&=&\sum_{T=0}^{\infty}\int_{0}^{\infty}\rmd A\; \rme^{-pA}Z_N(A,T)\nonumber\\
&=&\int_{0}^{\infty}\rmd A\;\rme^{-pA}Z_N(A)=[g(p,z=1)]^N,
\label{Z_mgf_laplace_z1}
\end{eqnarray}
where $Z_N(A)$ is given by
\begin{equation}
Z_N(A)=\int_{0}^{\infty}\rmd a_1\dots \int_{0}^{\infty}\rmd a_N \prod_{i=1}^N f_a(a_i)\, \delta\left(\sum_{j=1}^{N}a_{j}-A\right)
\end{equation}
and $f_a(a)= \sum_{\tau=1}^{\infty} f(a, \tau)$ is the (marginal) probability density of the area under an excursion. Setting $z=1$ in the expression (\ref{transform}) for $g(p,z)$ gives
\begin{equation}
g(p,z=1)=\frac{J_{2/p}(2/p)}{J_{2/p-1}(2/p)}=\frac{1}{1+\frac{J'_{2/p}(2/p)}{J_{2/p}(2/p)}},
\label{transform_z1}
\end{equation}
where in the last step we used the relation 
\begin{equation}
J_{\nu-1}(u)=\frac{\nu}{u}J_{\nu}(u)+J'_{\nu}(u).
\label{bessel_identity.1}
\end{equation}
We can apply the following asymptotic expansion of $J_{\nu}'(\nu)$ and $J_{\nu}(\nu)$ when $\nu\rightarrow\infty$ \cite{Olver54},
\begin{eqnarray}
J'_{\nu}&\sim& -\frac{2^{2/3}\textrm{Ai}'(0)}{\nu^{2/3}}+\Or\left(\frac{1}{\nu^{4/3}}\right)\qquad \nu\rightarrow\infty,\nonumber\\
J_{\nu}&\sim& \frac{2^{1/3}\textrm{Ai}(0)}{\nu^{1/3}}+\Or\left(\frac{1}{\nu}\right),\qquad \nu\rightarrow\infty,
\end{eqnarray}
where $\textrm{Ai(x)}$ and $\textrm{Ai}'(x)$ are the Airy function and its first derivative, respectively, which upon inserting into (\ref{transform_z1}) finally yields
\begin{eqnarray}
g(p,z=1)&=& 1+\frac{\textrm{Ai}'(0)}{\textrm{Ai}(0)}p^{1/3}+\Or(p^{2/3})\nonumber\\
&=& 1-\frac{3^{1/3}\Gamma(2/3)}{\Gamma(1/3)} p^{1/3}+\Or(p^{2/3}),\qquad p\rightarrow 0.
\end{eqnarray}
It is well known that the existence of non-integer powers in the moment-generating function around $p=0$ means that the corresponding probability distribution has power-law tails (see, for example, \cite{regular-variation}). Omitting here the details, the tail of $f(a)$ when $a$ is large is given by 
\begin{equation}
f_a(a)\sim\frac{1}{3^{2/3}\Gamma(1/3)}\frac{1}{a^{4/3}},\qquad 
a\rightarrow\infty.
\end{equation}
A similar expression has been obtained for the Brownian motion in Ref. \cite{KM05}. As before, the exponent $4/3$ is less than $2$ and hence, there is no condensation  transition.

\subsection{Condensation transition when both local time and total area are fixed}
\label{condensation-transition}

When both $l_T$ and $A_T$ are fixed, we have to study the full partition function $Z_N(A,T)$ which can be obtained by inverting $\mathcal{Z}_N(p,z)$ in (\ref{Z_mgf_laplace})
\begin{equation}
Z_N(A,T)=\int_{c-i\infty}^{c+i\infty}\frac{\rmd p}{2\pi i}\;\rme^{pA}\oint_{\gamma}\frac{\rmd z}{2\pi i}\frac{[g(p,z)]^N}{z^{T+1}},
\end{equation} 
where $\gamma$ is a contour around $z=0$ where $g(p,z)$ is analytic and $c$ is a real number such that $c$ is greater than the real part of all the singularities of the integrand.
Setting
\begin{equation}
T=\mu N, \qquad A=\sigma N,
\end{equation}
the partition function $Z_N(T,A)$ can be written as
\begin{equation}
Z_N(A,T)=\int_{c-i\infty}^{c+i\infty}\frac{\rmd p}{2\pi i}\oint_{\gamma}\frac{\rmd z}{2\pi i}\,\frac{1}{z}\,\rme^{N\,h(p,z)},
\label{Z}
\end{equation} 
where $h(p,z)$ is given by
\begin{equation}
h(p,z)=\textrm{ln}g(p,z)-\mu \textrm{ln}z+p\sigma.
\end{equation}

The standard technique to evaluate the integral  (\ref{Z}) for large $N$ is to use the saddle-point method, which amounts to solving the following saddle-point equations
\numparts
\begin{eqnarray}
\mu&=& z\frac{\partial}{\partial z}\textrm{ln}g(p,z)\label{saddle_point_eq1}\\
\sigma&=&-\frac{\partial}{\partial p}\textrm{ln}g(p,z)\label{saddle_point_eq2}.
\end{eqnarray}
\endnumparts

In general, if both saddle-point equations admit a solution denoted by $z=z_0$ and $p=p_0$ then $Z_N(\sigma N,\mu N)$ can be found by the saddle-point method yielding $Z_N(\sigma N,\mu N)\sim\textrm{exp}[Nh(p_0,z_0)]$ for large $N$. On the other hand, if one or both saddle-points have no solution, then the saddle-point method is no longer applicable, which typically signals a condensation transition. We show below that the equation (\ref{saddle_point_eq1}) has a  solution for any $\mu>1$, while the equation (\ref{saddle_point_eq2}) has a solution only for $0<\sigma<\sigma_c(\mu)$, where the transition point $\sigma_c(\mu)$ is given by  (\ref{sigmac}).

To analyse the saddle-point equations, it proves convenient to consider an auxiliary probability density $\omega(a,\tau)$ 
\begin{equation}
\omega(a,\tau; p, z)=\frac{f(a,\tau)\,z^{\tau}\, \rme^{-pa}}{g(p,z)}\;.
\label{omega}
\end{equation}
This auxiliary joint distribution, whose arguments are $a$ and $\tau$ and which is parametrised by $p$ and $z$, indeed corresponds to the grand canonical ensemble when $p$ and $z$ are chosen appropriately. The first saddle-point equation (\ref{saddle_point_eq1}) is then simply
\begin{equation}
\mu=\mathbb{E}_{\omega}[\tau],
\label{saddle_point_eq1_2}
\end{equation}
where the average is taken with respect to the probability density $\omega$. In \ref{appendix_c} we show that the function $\mathbb{E}_{\omega}[\tau](p,z)$ for fixed $p$ increases monotonically from $1$ to $\infty$ for $0\leq z<(pj_{2/p-1,1}/2)^2$, where $j_{\nu,k}$ is the $k$-th zero of the Bessel function $J_\nu$. This means that equation (\ref{saddle_point_eq1_2}) can be solved for any $\mu>1$ and we denote its solution by $z_0(p,\mu)$.

Using the probability density $\omega(a,\tau; p,z)$, we can write the second saddle-point equation (\ref{saddle_point_eq2}) as
\begin{equation}
\sigma=\mathbb{E}_{\omega}[a].
\end{equation}
In \ref{appendix_c} we show that $\mathbb{E}_{\omega}[a](p,z_0(p,\mu))$ is monotonically decreasing in $p$ and decays to $0$ when $p\rightarrow\infty$. 

To summarise,  saddle-point equation (\ref{saddle_point_eq1}) can always be satisfied for $1< \mu < \infty$ whereas  saddle-point equation (\ref{saddle_point_eq2}) can only be satisfied for sufficiently small $\sigma$. The maximal value of $\sigma$,  which we denote  $\sigma_c$, for which equation (\ref{saddle_point_eq2}) can be satisfied is given by taking $p\to 0$ in the right hand side and replacing $z$ by the value $z_0$ which solves equation (\ref{saddle_point_eq1}) as $p\to 0$:
\begin{equation}
\sigma_c=\mathbb{E}_{\omega}[a](0,z_0(p=0,\mu)). 
\end{equation}

To find $\sigma_c$, we thus need to analyse (\ref{transform}) in the $p\to 0$ limit. We use the following asymptotic expansions of the Bessel function $J_\nu(\nu z^{1/2})$ and its derivative $J_{\nu}'(\nu z^{1/2})$ when $\nu\rightarrow\infty$ \cite{Olver54}, (in our case we will identify $\nu =2/p$)
\numparts
\begin{eqnarray}
J_{\nu}(\nu z^{1/2})&\simeq&\left(\frac{4\zeta}{1-z}\right)^{1/4}\left[\frac{\textrm{Ai}(\nu^{2/3}\zeta)}{\nu^{1/3}}+\Or\left(\nu^{-5/3}\right)\right],\label{bessel1}\\
J_{\nu}'(\nu z^{1/2})&\simeq&-\frac{2}{z^{1/2}}\left(\frac{4\zeta}{1-z}\right)^{-1/4}\left[\frac{\textrm{Ai}'(\nu^{2/3}\zeta)}{\nu^{2/3}}+\Or\left(\nu^{-4/3}\right)\right]\label{bessel2},
\end{eqnarray}
\endnumparts
\noindent where $\textrm{Ai}(x)$ is the Airy function and $\zeta\equiv \zeta(z)$ is given by
\begin{equation}
\frac{2}{3}\, \zeta^{3/2}= \ln\left[\frac{1+\sqrt{1-z}}{\sqrt{z}}\right]-\sqrt{1-z}
\quad 0\le z \le 1\;.
\label{zeta.2}
\end{equation}
Using the identity (\ref{bessel_identity.1}) we can rewrite $g(p,z)$ in (\ref{transform}) as
\begin{equation}
g(p,z)=z\left[1+z^{1/2}\frac{J'_{2/p}(2z^{1/2}/p)}{J_{2/p}(2z^{1/2}/p)}\right]^{-1}.
\label{g-small-p}
\end{equation}
Combining (\ref{bessel1}), (\ref{bessel2}) and (\ref{g-small-p}) yields
\begin{equation}
g(p,z)=\frac{z}{1-(1-z)^{1/2}L(2^{2/3}p^{-2/3}\zeta)},\qquad p\rightarrow 0,
\label{g_nu_large}
\end{equation}
and the function $L(x)$, appearing in (\ref{g_nu_large}) is given by
\begin{equation}
L(x)=\frac{\textrm{Ai}'(x)}{x^{1/2}\textrm{Ai}(x)}.
\label{Ldef}
\end{equation}

Since we are considering $p\to 0$ and therefore large argument $x$ of the function $L(x)$, we can further make use of the following expansion of $L(x)$ for large, real $x$
\begin{equation}
L(x)=\frac{\textrm{Ai}'(x)}{x^{1/2}\textrm{Ai}(x)}=-1-\frac{1}{4x^{3/2}}+\Or(x^{-3}),\qquad x\rightarrow\infty.
\label{Kseries}
\end{equation}
Inserting (\ref{Kseries}) into (\ref{g_nu_large}), one obtains, for small $p >0$ and fixed $z$, the leading expression
\begin{equation}
g(p,z)\approx \frac{z}{1+\sqrt{1-z}+ \frac{p}{8}\, \sqrt{1-z}\, \zeta^{-3/2}}\, .
\label{gpzp0}
\end{equation}
Incidentally, we note that at $p=0$ one recovers the exact universal Sparre Andersen expression as in Eq. (\ref{gz0.0})
\begin{equation}
g(0,z)= \frac{z}{1+\sqrt{1-z}}= 1-\sqrt{1-z} \, .
\label{gz0.1}
\end{equation}
Finally, taking partial derivatives with respect to $z$ and $p$ gives
the following expressions for $z_0(0;\mu)$ and $\sigma_c(\mu)$,
\numparts
\begin{eqnarray}
z_0(0,\mu)&=& 1-\frac{1}{(2\mu-1)^2}\label{z00}\\
\sigma_c(\mu)&=&\frac{1}{12\mu\left[\textrm{ln}\left(\frac{\mu}{\mu-1}\right)-\frac{2}{2\mu-1}\right]}\label{sigmac}.
\end{eqnarray}
\endnumparts

\begin{figure}[hbt]
	\centering\includegraphics[width=10cm]{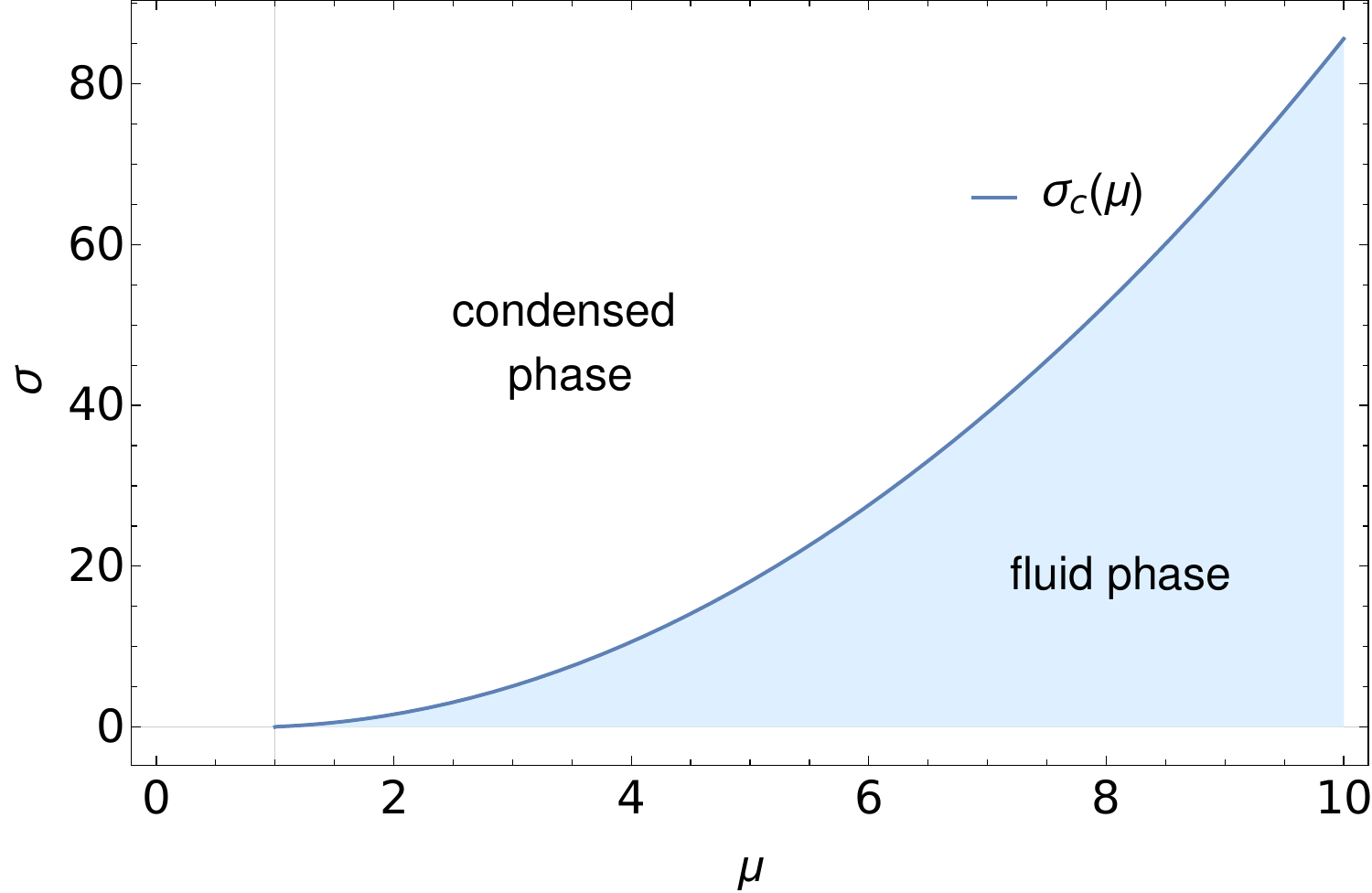}
	\caption{Phase diagram in the $\mu-\sigma$ plane consisting of the `fluid' phase and `condensed' phase for $0<\sigma<\sigma_c$ and $\sigma>\sigma_c$, respectively.}
	\label{fig3}
\end{figure}

The exact expression (\ref{sigmac}) for the critical line $\sigma_c(\mu)$, is the main result of this section. Its significance is that the saddle-point equation (\ref{saddle_point_eq2}) can be solved only for $0<\sigma<\sigma_c$ (the `fluid' phase), but has no solution for $\sigma>\sigma_c$ (the `condensed' phase). The phase diagram in the $\mu-\sigma$ plane with $\sigma_c(\mu)$ separating these two regimes is presented in Figure \ref{fig3}.

\subsection{Nature of the condensate}

To understand how the condensate arises in the system, let us write again the \emph{microcanonical} partition function $Z_{N}(A,T)$ 
\begin{equation}
\fl \quad Z_{N}(A,T)=\int_{0}^{\infty}\rmd a_1\dots\rmd a_N\sum_{\{\tau_i\}}\prod_{i=1}^{N}f(a_i,\tau_i)\delta\left(\sum_{k=1}^{N}\tau_{k}-T\right)\delta\left(\sum_{j=1}^{N}a_j-A\right),
\label{partition-micorcan}
\end{equation}
which describes the system of $N$ random walk excursions with fixed total area $A$ and total duration $T$. 

For $A=\sigma N$ and $T=\mu N$, the \emph{grand canonical} partition function is given by 
\begin{equation}
\fl \qquad \mathcal{Z}_{N}(p,z)=\int_{0}^{\infty}\rmd a_1\dots\rmd a_N\sum_{\{\tau_i\}}\prod_{i=1}^{N}f(a_i,\tau_i)\rme^{-pa_i}z^{\tau_i}=[g(p,z)]^N,
\end{equation}
where $z$ and $p$ are found by solving the saddle-point equations (\ref{saddle_point_eq1}) and (\ref{saddle_point_eq2}), respectively. In the previous Section we showed that the microcanonical and grand canonical ensembles are not equivalent when $\sigma>\sigma_c$, because we cannot find $p=p_0$ that solves the second saddle-point equation (\ref{saddle_point_eq2}). However, since we can always find the solution to the first saddle point equation (\ref{saddle_point_eq1}), we can replace the microcanonical part that enforces $l_T=N$ by the corresponding canonical one. The resulting partition function, 
\begin{equation}
\fl \qquad \mathcal{Y}_{N}(A,z)=\int_{0}^{\infty}\rmd a_1\dots\rmd a_N\left[\prod_{i=1}^{N}\sum_{\{\tau_i\}}f(a_i,\tau_i)z^{\tau_i}\right]\delta\left(\sum_{j=1}^{N}a_j-A\right),
\end{equation}
can be considered as a \emph{mixed} canonical-microcanonical partition function: canonical with respect to $l_T$ and microcanonical with respect to $A_T$ (see Section 5 in Ref. \cite{Ellis00} for a rigorous analysis).

For $\sigma>\sigma_c$, we know that the mixed canonical-microcanonical ensemble is equivalent to the microcanonical one for $z=z_0(p=0,\mu)=1-1/(2\mu-1)^2$, since that is the solution to the first saddle-point equation (\ref{saddle_point_eq1}). To calculate $\mathcal{Y}_N(A,z)$, we consider the marginal of $\omega(a,\tau; p, z)$  (\ref{omega}), evaluated at $z=z_0(p=0,\mu)$ and $p=0$
\begin{equation}
\omega(a)=\sum_{\tau=1}^{\infty}\omega(a,\tau; 0, z_0)= 
\frac{1}{g(0,z_0)}\, \sum_{\tau=1}^{\infty} f(a,\tau)\, z_0^{\tau} = 
\frac{\widetilde f(a,z_0)}{g(0,z_0)},
\label{omega-marginal}
\end{equation}
where we have used the definition of $\widetilde f(a,z)$ in (\ref{gtilde}) and $g(0,z)=1-\sqrt{1-z}$ from (\ref{gz0.1}).

The distribution $\omega(a)$ in (\ref{omega-marginal}) is thus the effective single-excursion area distribution within the mixed canonical-microcanonical ensemble.
It is straightforward to establish the following properties of $\omega(a)$:
\begin{equation}
\int_{0}^{\infty}\rmd a \;\omega(a)=1,\qquad \int_{0}^{\infty}\rmd a\; a\omega(a)=\sigma_c.
\end{equation}

Using $\omega(a)$, the mixed partition function $\mathcal{Y}_N(A,z_0)$ can be written as
\begin{equation}
\mathcal{Y}_{N}(A,z_0)=[g(0,z_0)]^N P\left(\sum_{i=1}^{N}a_i=A\right),
\end{equation}
where $P$ is the probability density of the sum of iid random variables $a_i$ with  common probability density $\omega(a_i)$,
\begin{equation}
P\left(\sum_{i=1}^{N}a_i=A\right)=\int_{0}^{\infty}\rmd a_1,\dots \rmd a_N \prod_{i=1}^{N}\omega(a_i)\delta\left(\sum_{j=1}^{N}a_j-A\right).
\label{Pa}
\end{equation}
The condensation thus arises because the canonical probability density $\omega(a_i)$ for the area of a single excursion, which emerges by conditioning random walks to fixed local time, becomes heavy-tailed. We explicitly calculate this heavy-tailed behaviour and show it to be  stretched exponential in section \ref{omega_tail}, equation (\ref{omegaasymp}). The second constraint on the total area $\sum_{i=1}^{N}a_i=\sigma N$ then forces the sum to have a large deviation, for which the condensation phenomenon is known to occur in which one random variable $a_i$ has a macroscopic size of $(\sigma-\sigma_c)N$.

This condensation mechanism is very similar to the one we previously studied in Ref. \cite{SNEM14a,SNEM14b}.  There the partition function of random variables $m_i$, $i=1,\dots,L$ was given by
\begin{equation}
\fl \quad Z_{L}(V,M)=\int_{0}^{\infty}\rmd m_1\dots m_L\prod_{i=1}^{L}f(m_i)\delta\left(\sum_{j=1}^{L}m_j-M\right)\delta\left(\sum_{j=1}^{L}m_{j}^{1/q}-V\right),
\label{ZVM}
\end{equation}
where $f(m_i)=r\rme^{-rm}$ and $0<q<1$ is a parameter appearing in the second delta function constraint in (\ref{ZVM}). In the condensed regime, the mixed canonical-microcanonical function for this problem reads
\begin{eqnarray}
\fl \qquad\mathcal{Y}_{L}(V,r)&=&\int_{0}^{\infty}\rmd m_1\dots\rmd m_N\left[\prod_{i=1}^{L}f(m_i)\rme^{-sm_i}\right]\delta\left(\sum_{j=1}^{N}m_{j}^{1/q}-V\right)\nonumber\\
\fl \qquad &=& \int_{0}^{\infty}\rmd v_1\dots\rmd v_N\prod_{i=1}^{L}\left[q v_{i}^{q-1}f(v_{i}^{q})\rme^{-sv_{i}^{q}}\right]\delta\left(\sum_{j=1}^{N}v_{j}-V\right),
\end{eqnarray}
where in the second line we made a change of variables $v_i=m_{i}^{1/q}$. The probability density of $v_i$ now has a stretched exponential heavy tail $\textrm{exp}(-(s+r)v^q)$, $0<q<1$, which is at the origin of this constraint-driven condensation.

\subsection{Asymptotic behaviour of \texorpdfstring{$\omega(a)$}{omega(a)}}
\label{omega_tail}

From Eq. (\ref{omega-marginal}), we have
\begin{equation}
\omega(a) = \frac{\widetilde f(a, z_0)}{g(0, z_0)}
\label{omegag}
\end{equation}
where $z_0$ is the solution (\ref{z00}). In order to compute the large $a$ behaviour of $\omega(a)$, we need to then compute ${\widetilde f(a,z)}$ from (\ref{gtilde}), by inverting the Laplace transform (with respect to $p$) of $g(p,z)$ given explicitly in (\ref{transform}), and finally replace $z$ by $z_0$.

\begin{figure}[hbt]
	\centering\includegraphics[width=10cm]{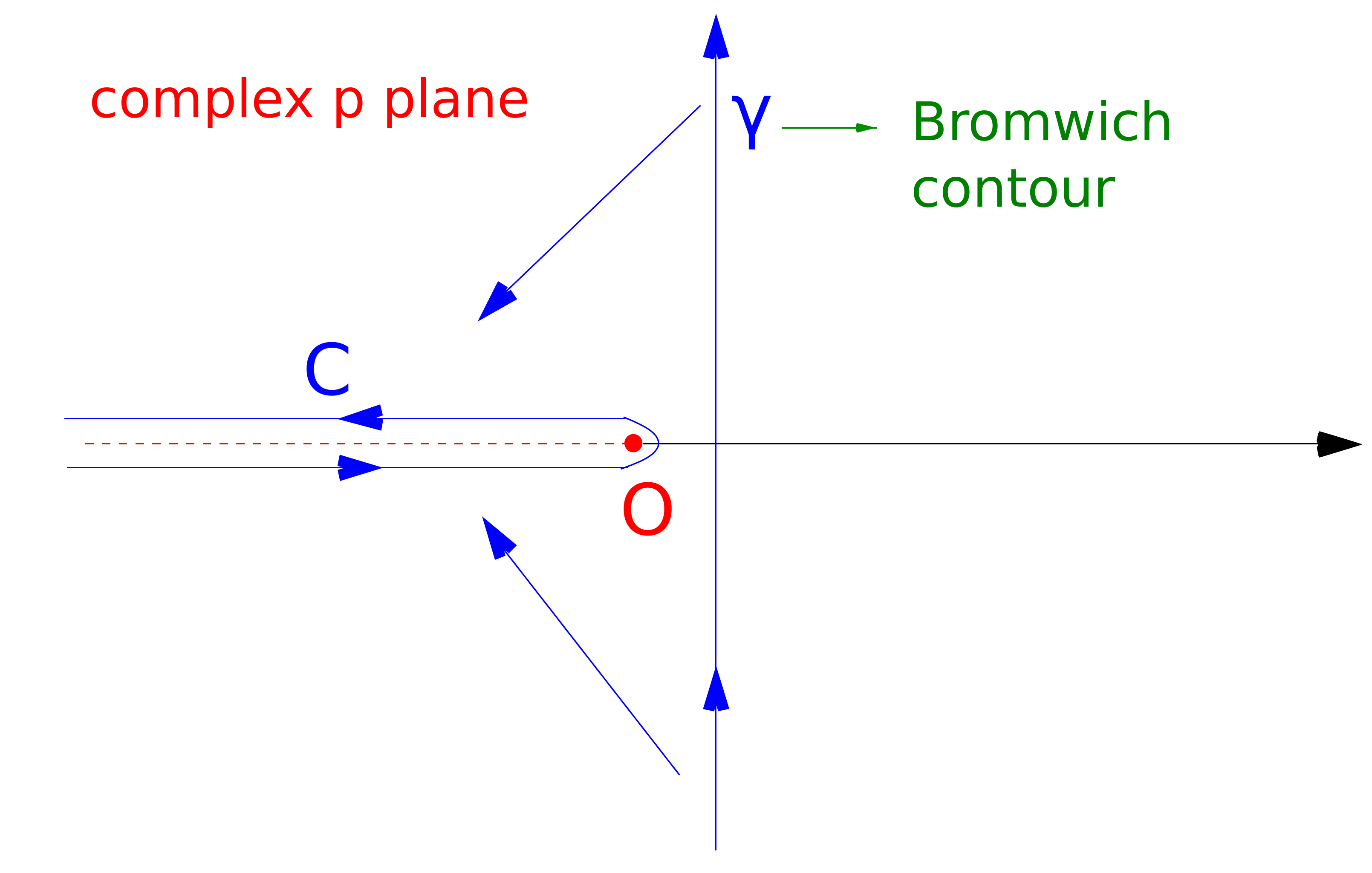}
	\caption{The branch cut in the complex $p$ plane is along the negative real $p$ axis (shown by dashed red lines) with the origin $O$ (red filled circle) as the branch point. The Bromwich contour $\gamma$ is deformed to the contour $C$ around the negative real $p$ axis.}
	\label{fig4}
\end{figure}

Exact computation of $\omega(a)$ via inverting this Laplace transform for arbitrary $a$ is rather hard. However, one can make progress for large $a$, by exploiting the singularity structure of $g(p,z)$ in the complex $p$ plane. We note that $g(p,z)$ has a branch cut for $\textrm{Re}[p]\le 0$ with $p=0$ being a branch point. Therefore we may deform the Bromwich contour in the complex $p$ plane (\ref{gtilde}) into a contour $C$ which extends from $\infty\, {\rm e}^{-i(\pi-\delta)}$, around the origin anti clockwise and back to $-\infty$ above the branch cut along $\infty\, {\rm e}^{+i(\pi-\delta)}$ where $\delta$ is some infinitesimal real number (see Fig. \ref{fig4}). In the following we will neglect $\delta$ to lighten the notation. We expect the inversion integral to be dominated for large $a$ by contributions near to the branch point at $p=0$, therefore we develop an expansion for small $p$, along the deformed contour $C$.

We first consider the function $L(x)$ (\ref{Ldef}) where $x= 2^{2/3}p^{-2/3}\zeta$. In the case $p=\epsilon{\rm e}^{\pm\pi i}$ where $\epsilon \to 0$, we need to  consider $L(y {\rm e}^{\mp 2\pi i/3})$ as $y\to \infty$ where
\begin{equation}
y= 2^{2/3}\epsilon^{-2/3}\zeta \;.
\label{ydef}
\end{equation}
We use the following connection formulae \cite{Olver54}
\begin{eqnarray}
\textrm{Ai} (y {\rm e}^{\pm 2\pi i/3})
&=& \frac{1}{2} {\rm e}^{\pm \pi i/3} \left[
\textrm{Ai} (y) \mp i \textrm{Bi} (y)\right] \label{Aicon}\\
\textrm{Ai}' (y {\rm e}^{\pm 2\pi i/3})
&=& \frac{1}{2} {\rm e}^{\mp \pi i/3} \left[
\textrm{Ai}' (y) \mp i \textrm{Bi}' (y)\right] \label{Aipcon}
\end{eqnarray}
Using (\ref{Aicon},\ref{Aipcon}) and the definition (\ref{Ldef}) we may write
\begin{eqnarray}
L(y {\rm e}^{\mp 2\pi i/3})&=&
-\frac{1}{y^{1/2}} \frac{
\left[ \textrm{Ai}' (y) \pm i \textrm{Bi}' (y)\right] }
{
\left[ \textrm{Ai} (y) \pm i \textrm{Bi} (y)\right] }\\
&=&
-\frac{1}{y^{1/2}}
\frac{\textrm{Bi}' (y)}{ \textrm{Bi} (y)}
\left[
1 \mp i\frac{\textrm{Ai}' (y)}{ \textrm{Bi}' (y)}
\pm  i\frac{\textrm{Ai}(y)}{ \textrm{Bi}(y)} + \cdots \right]
\label{Lexp}
\end{eqnarray}
Now although $\displaystyle \frac{1}{y^{1/2}} \frac{\textrm{Bi}' (y)}{ \textrm{Bi} (y)}$ is non-analytic in $y$, by virtue of the relation (\ref{ydef}) it is an analytic function of $\epsilon$. This function  which tends to $-1$ as $\epsilon \to 0$ ($y\to \infty$), thus the leading non-analytic  terms in (\ref{Lexp}) come from the two ratios $\textrm{Ai}' (y)/\textrm{Bi}'(y)$ and $\textrm{Ai} (y)/\textrm{Bi}(y)$. We use the  asymptotic behaviours for large $y$ (small $\epsilon$)
\begin{eqnarray}
\textrm{Ai}(y) &\sim& \frac{1}{2\pi^{1/2}}\, y^{-1/4}\, {\rm e}^{-u} \\
\textrm{Ai}'(y) &\sim& -\frac{1}{2\pi^{1/2}}\, y^{1/4}\, {\rm e}^{-u} 
\\
\textrm{Bi}(y) &\sim& \frac{1}{\pi^{1/2}}\, y^{-1/4}\, {\rm e}^{u} \\
\textrm{Bi}'(y) &\sim& -\frac{1}{\pi^{1/2}}\, y^{1/4}\, {\rm e}^{u} 
\end{eqnarray}
where
\begin{equation}
u = \frac{2}{3} y^{3/2}
\end{equation}
to obtain the leading non-analytic behaviour of $L$ as $y \to \infty$
\begin{equation}
L(y {\rm e}^{\mp 2\pi i/3})\sim
-1 \mp i \exp \left(-\frac{4 y^{3/2}}{3}\right)\;.
\end{equation}

We use this expression and equations (\ref{ydef}) and (\ref{g_nu_large}), to evaluate the leading nonanalytic term in the  expansion of $g(p,z)$, for $p=\epsilon{\rm e}^{\pm\pi i}$ where $\epsilon \to 0$,
\begin{equation}
\frac{g(\epsilon{\rm e}^{\pm\pi i}, z)}{g(0,z)}
\sim
1 %\frac{z}{(1+(1-z^{1/2})}
\mp i \frac{(1-z)^{1/2}}{(1+(1-z)^{1/2})}
\exp \left(-\frac{b}{\epsilon}\right)
\end{equation}
where
\begin{equation}
b = \frac{8 \zeta^{3/2}}{3}
\label{bdef}
\end{equation}
and $\zeta$ is defined as before in (\ref{zeta.2}). Using the deformed contour $C$ we now obtain equal contributions from pieces above and below the branch cut and  we obtain
\begin{equation}
\frac{\widetilde f(a,z)}{g(0,z)}  \sim  \frac{1}{\pi}
\frac{(1-z)^{1/2}}{(1+(1-z)^{1/2})}
\int_0^{\infty} \exp \left(-\frac{b}{\epsilon} -\epsilon a\right) {\rm d}\epsilon
\label{fint}
\end{equation}

Note that in (\ref{fint}) we may use the small $\epsilon$ behaviour even though
the limit of integration extends to  $\infty$, because for large $a$ the integral is dominated by small $\epsilon$. In fact, we can evaluate (\ref{fint}) by a straightforward application of the saddle point method to find the integral is dominated by contributions around  $\epsilon = (b/a)^{1/2}$ and we obtain for $a \gg 1$
\begin{equation}
\frac{\widetilde f(a, z)}{g(0, z)} \simeq  \frac{1}{\pi^{1/2}}
\frac{b^{1/4}}{a^{3/4}}
\frac{(1-z)^{1/2}}{(1+(1-z)^{1/2})}
\exp \left( -2 b^{1/2}a^{1/2}\right)
\label{fasymp}
\end{equation}

Finally we can insert the value of $z=z_0$ into (\ref{fasymp}) and use the definition (\ref{omega-marginal}) to find the asymptotic behaviour of the single-excursion area distribution:
\begin{equation}
\omega(a)
\simeq  \frac{1}{\pi^{1/2}}
\frac{b^{1/4}}{a^{3/4}}
\frac{1}{2\mu}
\exp \left( -2 b^{1/2}\,a^{1/2}\right)
\label{omegaasymp}
\end{equation}
This expression confirms a stretched exponential form which obeys the criterion for a condensation transition \cite{EMZ06}. Thus the effect of the joint constraints is to induce an effective single excursion area distribution which admits a condensation transition for the area constraint in (\ref{Pa}).

\section{Interaction-driven condensation in pair-factorised steady states}
\label{interaction-driven}

In this Section we make an explicit connection between reflected random walk paths conditioned on fixed local time and area and pair-factorised steady states in mass-transfer models which exhibit interaction-driven condensation.

\subsection{A short review of the interaction-driven condensation}

Interaction-driven condensation is a novel type of condensation that was reported first by Evans, Hanney and Majumdar in 2006 \cite{EHM06}. The condensation takes place in one-dimensional mass-transfer models in which discrete masses $m_i\geq 0$, $i=1,\dots,L$ are exchanged at rates that depend on their immediate environment, which mimics local interaction between the masses. 

Under certain conditions on the dynamics, the steady state can be found explicitly and takes the following pair-factorised form:
\begin{equation}
P(m_1,\dots,m_L)=\frac{1}{Z_{L}(M)}\prod_{i=1}^{L}g(m_{i},m_{i+1})\delta\left(\sum_{j=1}^{L}m_j-M\right)\;,
\label{pair-factorised-probability2}
\end{equation}
where the partition function $Z_L(M)$ is given by
\begin{equation}
Z_{L}(M)=\sum_{\{m_i\}}\prod_{i=1}^{L}g(m_{i},m_{i+1})\delta\left(\sum_{j=1}^{L}m_j-M\right).
\label{partition-pair}
\end{equation}
The weight function $g(m_i,m_{i+1})$ reads
\begin{equation}
g(m_i,m_{i+1})=\textrm{exp}\left[-J\vert m_{i+1}-m_{i}\vert+{1\over 2}U\delta_{m_i,0}+{1\over 2}U\delta_{m_{i+1},0}\right],
\label{g_EHM}
\end{equation}
where $J$ and $U$ are positive constants. For the choice of $g(m_{i},m_{i+1})$ in (\ref{g_EHM}), the condensation occurs for the density $\rho=M/L$ greater than the critical density $\rho_c$,

\begin{equation}
	\rho_c=\frac{1}{\rme^{2J}(1-\rme^{-U})^2-1}\;.
	\label{rhoc}
\end{equation}

This condensation differs from the standard one found in mass-transfer models with factorised steady states: here the condensate does not reside on a single site but extends over $\Or(L^{1/2})$ sites; for this reason, the phenomenon is also known as spatially-extended condensation \cite{WSJM09a,WSJM09b,ENJ14}.

\subsection{Mapping pair factorised steady states to the reflected random walk paths}

Inserting (\ref{g_EHM}) into (\ref{partition-pair}) gives 
\begin{equation}
\fl \qquad Z_{L}(M)=\sum_{\{m_i\geq 0\}}\left[\prod_{i=1}^{L}\rme^{-J\vert m_{i+1}-m_{i}\vert}\right]\rme^{U\sum_{j=1}^{L}\delta_{m_j,0}}\delta\left(\sum_{k=1}^{L}m_k-M\right).
\label{partition-pair2}
\end{equation}
The expression in square brackets is proportional to the path probability for the standard random walk,
\begin{equation}
\prod_{i=1}^{L}\rme^{-J\vert m_{i+1}-m_{i}\vert}=\left(\frac{\rme^{J}+1}{\rme^{J}-1}\right)^{L}\prod_{i=1}^{L}
\theta(m_{i+1}-m_i),
\label{discrete-rw-path}
\end{equation}
where $\theta(\eta)$ is the discrete double exponential jump distribution
\begin{equation}
\theta(\eta)=\left(\frac{\rme^{J}-1}{\rme^{J}+1}\right)\rme^{-J\vert \eta\vert},\qquad \sum_{\eta=-\infty}^{\infty}\theta(\eta)=1.
\end{equation}
However, since the summation in $Z_L(M)$ is over all non-negative $m_i$, $i=1,\dots,L$ we would like to rewrite	the above expression in terms of the reflected rather than the standard random walk. This is indeed possible for the double exponential jump distribution and the corresponding transition probability reads 
\begin{eqnarray}
\fl\qquad w(m_{i+1}\vert m_{i})&=& \delta_{m_{i+1},0}\left[\sum_{\eta=-\infty}^{-m_{i}}\theta(\eta)\right]+[1-\delta_{m_{i+1},0}]\theta(m_{i+1}-m_{i})\nonumber\\
&=& \theta(m_{i+1}-m_{i})\rme^{R\delta_{m_{i+1},0}},
\label{discrete-rrw-path}
\end{eqnarray}
where $R=J-\textrm{ln}(\rme^J-1)$. Inserting (\ref{discrete-rrw-path}) into (\ref{partition-pair2}) gives
\begin{equation}
\fl \qquad Z_{L}(M)=\sum_{\{m_i\geq 0\}}\prod_{i=1}^{L}w(m_{i+1}\vert m_{i})\rme^{(U-R)l_L}\delta\left(\sum_{k=1}^{L}m_k-M\right),
\label{partition-pair3}
\end{equation}
where $l_L=\sum_{j=1}^{L}\delta_{m_j,0}$. The only thing left to do is to interpret the exponential weight $\textrm{exp}[(U-R)l_L]$, which we do in terms of nonequilibrium path ensembles.

\subsection{Nonequilibrium path ensembles and their equivalence}

First, we note that the sum in the exponential in (\ref{partition-pair3}) is just the local time $l_L$
\begin{equation}
\rme^{(U-R)\sum_{i=1}^{L}\delta_{m_i,0}}=\rme^{(U-R)l_L}.
\label{potential-canonical}
\end{equation}
Second, we argue that the exponential factor $\textrm{exp}[(U-R)l_L]$ represents the driven path ensemble (also known  as $s$ ensemble for short) with respect to $l_L$, in contrast to the microcanonical or conditioned one in which $l_L$ is fixed. Such nonequilibrium paths extend the notion of statistical (thermodynamic) ensembles to dynamical trajectories of systems which are not necessarily in the thermal equilibrium \cite{CT15,SNE15}.

The driven or $s$ ensemble is described by the exponentially tilted path probability
\begin{equation}
P_s[m]=\frac{\left[\prod_{i=1}^{L}w(m_{i+1}\vert m_i)\right]\rme^{sl_L}}{\langle\rme^{sl_L}\rangle}.
\label{path-driven}
\end{equation}
On the other hand, the microcanonical ensemble conditioned on a large deviation $l_L=N=\lambda L$ is described by the path probability
\begin{equation}
P[m\vert l_L=\lambda L]=\frac{\prod_{i=1}^{L}w(m_{i+1}\vert m_i)\delta(l_T-\lambda L)}{\langle\delta(l_T-\lambda L)\rangle}.
\label{path-conditioned}
\end{equation}
Here, the asymptotic equivalence means that
\begin{equation}
\lim_{L\rightarrow\infty}\frac{1}{L}\textrm{ln}\frac{P_s[m]}{P[m\vert l_L=\lambda L]}=0
\label{equivalence}
\end{equation}
almost everywhere with respect to $P_s[m]$ \cite{CT13,CT15}. The equivalence holds provided (a) the probability distribution $P(l_L=\lambda L)$ for the reflected random walk obeys a large deviations principle, 
\begin{equation}
P(l_L=\lambda N)\sim \rme^{-LI(\lambda)},\qquad L\rightarrow\infty
\end{equation}
and (b) the corresponding rate function $I(\lambda)$ is convex. In addition, if the rate function $I(\lambda)$ is differentiable, then the titling parameter $s$ is unique and is given by $s=I'(\lambda)$.

Assuming that the equivalence (\ref{equivalence}) holds, we can choose $\lambda$ such that $U-R=I'(\lambda)$. The partition function $Z_L(M)$ in (\ref{partition-pair3}) is then equivalent to the following microcanonical partition function
\begin{equation}
\fl \qquad Z_{L}(M,N)=\sum_{\{m_i\geq 0\}}\prod_{i=1}^{L}w(m_{i+1}\vert m_{i})\delta\left(\sum_{j=1}^{L}\delta_{m_j,0}-N\right)\delta\left(\sum_{k=1}^{L}m_k-M\right),
\label{partition-pair4}
\end{equation}
which is precisely the partition function (\ref{partition-function}), now for the discrete random variables $m_i$, for the reflected random walk conditioned on the area $A=M$ and local time $l_L=N$. 

This is our second main result: it shows that the pair-factorised steady state defined in (\ref{pair-factorised-probability2}), which involves correlated random variables $m_i$, $i=1,\dots,L$ conditioned on the fixed sum $\sum_{i}m_i=M$, can be instead understood in terms of reflected random walk paths conditioned on the number of returns $l_L=\sum_{i}\delta_{m_i,0}=N$ to the origin and total area $A_L=\sum_i m_i =M$. 

From there, one follows the procedure described in Section \ref{conditioned-random-walk} to reformulate random walk paths in terms of $N$ independent random walk excursions having area $a_j$ and duration $\tau_j$, $j=1\dots,N$. At this point the analysis is very similar to the constraint-driven condensation that we recently studied in Ref. \cite{SNEM14a,SNEM14b} for joint large deviations of sums $\sum_j m_j=M$ and $\sum_j m_{j}^{2}=V$, the only difference being that random variables $a_j$ and $\tau_j$ are correlated but not functionally dependent on each other. The mechanism of condensation is nevertheless the same: fixing the total duration $\sum_j\tau_j=T=\mu N$ is responsible for causing $a_j$ to become essentially heavy tailed, while fixing the sum $\sum_j a_j=M=\sigma N$ ensures that the sum represents a large deviation resulting in the standard condensation phenomenon for $\sigma>\sigma_c(\mu)$.

%===============================================================================%
% 5. CONCLUSION
%===============================================================================%
\section{Conclusions}
\label{conclusion}

In this work we have considered a discrete-time continuous-space reflected random walk under the constraints that the number of returns to the origin and the total area under the walk are fixed.  We have shown how condensation occurs through the imposition of extensive values (proportional to the walk duration) on the number of returns $l_T=\sum_{t}\delta_{x_t,0}=N$ to the origin and total area $A_T=\sum_t x_t =M$, whereas it would not occur if only the number of returns or the total area were constrained.

In order to study the condensation we derived the joint probability $f(a,\tau)$ of a reflected random walk excursion having area $a$ and returning to the origin for the first time after time $\tau$.  This distribution is of some interest in its own right and we have included some further properties of the distribution in \ref{appendix_b}.

We have also seen how `interaction-driven', spatially-extended condensation exhibited in pair-factorised steady states can also be understood in terms of reflected random walk paths.  It would be of interest to see if our approach can allow further properties of spatially-extended condensates to be analysed.

More generally, the problem of the random walk with constraints on local time and total area is an example of two correlated, but not functionally dependent random variables. The correlation is then sufficient to induce condensation in one of the random variable which would not otherwise exhibit condensation. In principle, conditions for condensation for random walk models with more general jump distributions should follow from the knowledge of the corresponding function $g(z,p)$. It would be of interest to confirm numerically (or if possible analytically) that the same scaling behaviour as established here, for example for the  single-excursion area distribution $\omega(a)$ in (\ref{omegaasymp}), holds  for more general random walk models.

Finally, we mention that in the context of mass-transfer models discussed in Section \ref{interaction-driven}, different choices of constraints may also lead to condensation behaviour and to different shape of the condensate \cite{WSJM09a,WSJM09b}. In the present work we have used as our main tool the renewal nature of the process which results from the local time constraint. However the results of Refs. \cite{WSJM09a,WSJM09b} suggest that an explicit renewal property is not strictly required for condensation. For example, constraining fractional moments of the walk forces the walk close enough to the origin that effectively one has a renewal picture.

%------------------------------------------------
% ACKNOWLEDGEMENTS
%------------------------------------------------
\ack
JSN and MRE would like to acknowledge funding from EPSRC under grant number EP/J007404/1. 
%
%---------------------------------------------------------
% A. APPENDIX: 
%---------------------------------------------------------
\appendix
\section{Calculation of the constant \texorpdfstring{$A(p,z)$}{A(p,z)}}
\label{appendix_a}

Inserting the solution (\ref{sol.1}) for   $G(x,p,z)$,
\begin{equation}
G(x,p,z)=A(p,z)\, \rme^{-px}\, J_{2/p}\left(2z^{1/2}\rme^{-px/2}/p\right),
\end{equation}
into the integral equation (\ref{Geq}) yields
\begin{eqnarray}
\fl \qquad A(p,z)J_{2/p}(2z^{1/2}/p) &=& zA(p,z)\int_{0}^{\infty}\rmd x' \frac{\rme^{-\vert x'\vert}}{2}J_{2/p}\left(2z^{1/2}\rme^{-px'/2}/p\right)\rme^{-px'}\nonumber\\
&&+z\int_{-\infty}^{0}\rmd x' \frac{\rme^{-\vert x'\vert}}{2}
\label{Geq2}
\end{eqnarray}
In the equation above we chose $x=0$ to simplify the calculation, since $A(p,z)$ does not depend on $x$. The last integral in the equation above equals $1/2$; to calculate the first integral we make the following change of variables
\begin{equation}
y=2 z^{1/2}\rme^{-px'/2}/p, \qquad dy=-z^{1/2}\rme^{-px'/2}dx'=(-p/2)y dx'.
\end{equation}
Introducing $c=2z^{1/2}/p$, the first integral then becomes
\begin{eqnarray}
\fl \int_{0}^{\infty}\rmd x'\rme^{-(p+1)x'}J_{2/p}\left(2z^{1/2}\rme^{-px'/2}/p\right)&=&
\frac{1}{z^{1/2}c^{2/p+1}}\int_{0}^{c}\rmd y\, y^{2/p+1}J_{2/p}(y)\nonumber\\
&=& \frac{1}{z^{1/2}c^{2/p+1}}\int_{0}^{c}\rmd y \frac{\rmd}{\rmd y}\left(y^{2/p+1}J_{2/p+1}(y)\right)\nonumber\\
&=& \frac{J_{2/p+1}(2z^{1/2}/p)}{z^{1/2}}.
\label{G_1_int}
\end{eqnarray}
Here in going from the first to the second line we use the following property of the Bessel function $J_{\nu}(y)$ for $\nu=2/p$,
\begin{equation}
\frac{\rmd}{\rmd y}\left[y^{\nu+1}J_{\nu+1}(y)\right]=y^{\nu+1}J_{\nu}(y).
\end{equation}
Inserting (\ref{G_1_int}) into (\ref{Geq2}) gives
\begin{equation}
\frac{A(p,z)z^{1/2}}{2}\left[\frac{2}{z^{1/2}}J_{2/p}-J_{2/p+1}\right]=\frac{z}{2}.
\end{equation}
The expression in square brackets is just $J_{2/p-1}(2z^{1/2}/p)$ so that the constant $A(p,z)$ is finally given by (\ref{Apz.1})
\begin{equation}
A(p,z)=\frac{z^{1/2}}{J_{2/p-1}(2z^{1/2}/p)}.
\end{equation}
%
%---------------------------------------------------------
% B. APPENDIX: 
%---------------------------------------------------------
\appendix
\setcounter{section}{1}
\section{Properties of \texorpdfstring{$f(a,\tau)$}{f(a,tau)} for the double exponential jump distribution}
\label{appendix_b}

\subsection{Explicit calculation of \texorpdfstring{$f(a,\tau)$}{f(a,tau)} for small \texorpdfstring{$\tau$}{tau}}

\begin{figure}[hbt]
	\centering\includegraphics[width=10cm]{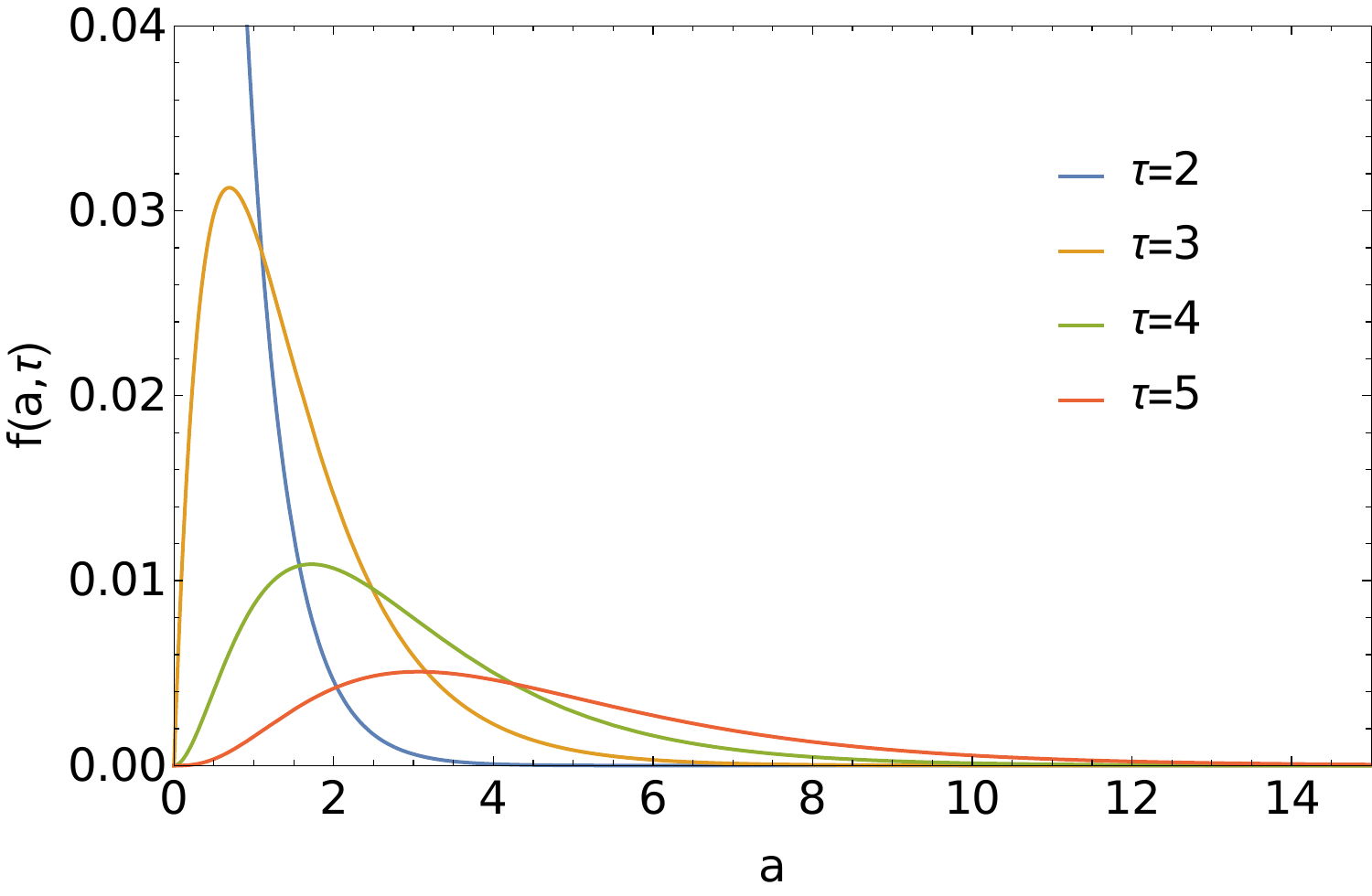}
	\caption{The joint probability $f(a,\tau)$ for $2\leq\tau\leq 5$ for the double exponential jump distribution.}
	\label{figA1}
\end{figure}

Let us recall the expression for the Laplace transform/moment-generating function $g(p,z)$
\begin{equation}
g(p,z)=\int_{0}^{\infty}\rmd a\;\rme^{-pa}\sum_{\tau=1}^{\infty}f(a,\tau)z^{\tau}=\frac{z^{1/2} J_{2/p}(2z^{1/2}/p)}{ J_{2/p-1}(2z^{1/2}/p)}.
\label{transform_2}
\end{equation}
Using the relation
\begin{equation}
J_{\nu}(u)=\frac{\nu-1}{u}J_{\nu-1}(u)-J'_{\nu-1}(u),
\end{equation}
we can write $g(p,z)$ as 
\begin{equation}
g(p,z)=1-\frac{p}{2}-z^{1/2}\frac{J'_{2/p-1}(2z^{1/2}/p)}{J_{2/p-1}(2z^{1/2}/p)}.
\label{transform_3}
\end{equation}
The Weierstrass factorisation for the Bessel function \cite{Kishore63} is given by
\begin{equation}
\label{Weierstrass}
J_{\nu}(u)=\frac{\left(u/2\right)^{\nu}}{\Gamma(\nu+1)}\prod_{m=1}^{\infty}\left(1-\frac{u^2}{j_{\nu,m}^{2}}\right),
\end{equation}
where $j_{\nu,m}$ is the $m$-th zero of the Bessel function $J_{\nu}(u)$. Using (\ref{Weierstrass}) we can write
\begin{eqnarray}
u\frac{J'_{\nu}(u)}{J_{\nu}(u)}&=& u\frac{\rmd}{\rmd u}\textrm{ln}J_{\nu}(u)=u\frac{\rmd}{\rmd u}\left[\nu \textrm{ln}u+\sum_{m=1}^{\infty}\textrm{ln}\left(1-\frac{u^2}{j_{\nu,m}^2}\right)\right]\nonumber\\
&=&\nu-2\sum_{m=1}^{\infty}\frac{u^2}{j_{\nu,m}^{2}-u^2}=\nu-2\sum_{k=1}^{\infty}u^{2k}\left(\sum_{m=1}^{\infty}\frac{1}{j_{\nu,m}^{2k}}\right)\nonumber\\
&\equiv&\nu-2\sum_{k=1}^{\infty}u^{2k}\sigma_{k}(\nu),
\label{derivative-log-bessel}
\end{eqnarray}
where the function $\sigma_{k}(\nu)$, which is known as the Rayleigh function \cite{Kishore63}, is given by
\begin{equation}
\label{Rayleigh}
\sigma_{k}(\nu)=\sum_{m=1}^{\infty}\frac{1}{j_{\nu,m}^{2k}}.
\end{equation}
Inserting (\ref{Rayleigh}) and (\ref{derivative-log-bessel}) into (\ref{transform_3}) gives
\begin{equation}
g(p,z)=\sum_{\tau=1}^{\infty}\frac{4^\tau}{p^{2\tau-1}}\sigma_{\tau}(2/p-1)z^\tau.
\label{g-series}
\end{equation}
From here it follows that the Laplace transform of $f(a,\tau)$ is given by
\begin{equation}
\int_{0}^{\infty}\rmd a\; \rme^{-pa}f(a,\tau)=\frac{4^\tau}{p^{2\tau-1}}\sigma_{\tau}(2/p-1).
\label{Laplace_f_2}
\end{equation}
The Rayleigh function $\sigma_{k}(\nu)$ is a rational function of $\nu$ and can be calculated recursively \cite{Kishore63} from
\begin{equation}
(\nu+k)\sigma_{k}(\nu)=\sum_{j=1}^{k-1}\sigma_{j}(\nu)\sigma_{k-j}(\nu),\quad \sigma_{1}(\nu)=\frac{1}{4(\nu+1)}.
\end{equation}
For example, for $k=1,\dots,7$ we get
\begin{eqnarray*}
&&\sigma_{1}(\nu)=\frac{1}{4(\nu+1)},\nonumber\\
&&\sigma_{2}(\nu)=\frac{1}{4^2(\nu+2)(\nu+1)^2},\nonumber\\
&&\sigma_{3}(\nu)=\frac{2}{4^3(\nu+3)(\nu+2)(\nu+1)^3},\nonumber\\ 
&&\sigma_{4}(\nu)=\frac{5\nu+11}{4^4(\nu+4)(\nu+3)(\nu+2)^2(\nu+1)^4},\nonumber\\
&&\sigma_{5}(\nu)=\frac{14\nu+38}{4^5(\nu+5)(\nu+4)(\nu+3)(\nu+2)^2(\nu+1)^5},\nonumber\\
&&\sigma_{6}(\nu)=\frac{42\nu^3+362\nu^2+1026\nu+946}{4^6(\nu+6)(\nu+5)(\nu+4)(\nu+3)^2(\nu+2)^3(\nu+1)^6},\nonumber\\
&&\sigma_{7}(\nu)=\frac{132\nu^3+1316\nu^2+4324\nu+3580}{4^7(\nu+7)(\nu+6)(\nu+5)(\nu+4)(\nu+3)^2(\nu+2)^3(\nu+1)^7}.
\end{eqnarray*}
For small $\tau$ we can invert the Laplace transform explicitly using partial fractions. The function $f(a,\tau)$ for $1\leq\tau\leq 5$ is listed below and plotted in Figure \ref{figA1}. The calculation for higher $\tau$ using partial fractions soon becomes cumbersome, but can be done e.g. using Mathematica software.
\begin{eqnarray*}
&& \fl\quad f(a,1)=\frac{\delta(a)}{2},\nonumber\\
&& \fl\quad f(a,2)=\frac{\rme^{-2a}}{4},\nonumber\\
&& \fl\quad f(a,3)=\frac{\rme^{-a}-\rme^{-2a}}{8},\nonumber\\
&& \fl\quad f(a,4)=\frac{27\rme^{-2a/3}-32\rme^{-a}-(4a-5)\rme^{-2a}}{256},\nonumber\\
&& \fl\quad f(a,5)=\frac{512\rme^{-a/2}-729\rme^{-2a/3}+144\rme^{-a}+(60a-73)\rme^{-2a}}{4608}.
\end{eqnarray*}

\subsection{Scaling limit of \texorpdfstring{$f(a,\tau)$}{f(a,tau)}}

Here we show another interesting result, the scaling limit of $f(a,\tau)$ when $a,\tau\rightarrow\infty$ with $a\tau^{-3/2}$ fixed. 

We start from the expression for $g(p,z)$ in (\ref{g-series}) with the Rayleigh function $\sigma_\tau(\nu)$ defined in (\ref{Rayleigh}). We then use asymptotic expansion for the zeroes of the Bessel function \cite{Olver51}
\begin{equation}
\label{Bessel0_asym}
j_{\nu,m}=\nu+\alpha_m\frac{\nu^{1/3}}{2^{1/3}},\qquad \nu\rightarrow\infty,
\end{equation}
where $-\alpha_m$ are the zeroes of the Airy function. Inserting (\ref{Bessel0_asym}) in (\ref{Laplace_f_2}) and using (\ref{Rayleigh}) gives 
\begin{eqnarray}
\int_{0}^{\infty}\rmd a\; \rme^{-pa}f(a,\tau)&=&\frac{4^\tau}{p^{2\tau-1}}\sum_{m=1}^{\infty}\frac{1}{j_{2/p-1,m}^{2\tau}}\nonumber\\
&\approx& p\sum_{m=1}^{\infty}\left(1+\alpha_m p^{2/3}/2\right)^{2\tau}\nonumber\\
&\approx& p\sum_{m=1}^{\infty}\rme^{-\alpha_m p^{2/3}\tau}
\end{eqnarray}
Let us now introduce $s=2^{1/2}p/\tau^{3/2}$ and $u=a/(2^{1/2}\tau^{3/2})$ so that
\begin{equation}
\int_{0}^{\infty}\rmd u\; \rme^{-su}f(u,\tau)=\frac{1}{2\sqrt{2\pi}\tau^3}\left[\sqrt{2\pi}s\sum_{m=1}^{\infty}\rme^{-\alpha_m 2^{-1/3}s^{2/3}}\right].
\end{equation}
The expression in the square brackets is Laplace transform of the Airy probability density $f_{\textrm{Airy}}$ \cite{MC05}. We thus conclude that $f(a,\tau)$ behaves for large $a$ and $\tau$ as
\begin{equation}
\label{f_scaling}
f(a,\tau)\approx \frac{1}{2\sqrt{2\pi}\tau^3}f_{\textrm{Airy}}\left(\frac{a}{2^{1/2}\tau^{3/2}}\right),\qquad \tau\rightarrow\infty,
\end{equation}
where the Airy probability density $f_{\textrm{Airy}}$ is defined by
\begin{equation}
f_{\textrm{Airy}}(x)=\frac{2^{13/6}}{3^{3/2}x^{10/3}}\sum_{m=1}^{\infty}\alpha_{m}^{2}\rme^{-2\alpha_{m}^{3}/(27x^2)}U\left(-\frac{5}{6},\frac{4}{3},\frac{2\alpha_{m}^{3}}{27x^2}\right)
\end{equation}
and $U$ is a confluent hypergeometric function.

It is further interesting to consider the conditional probability density $f(a\vert \tau)$ defined by
\begin{equation}
f(a\vert \tau)=\frac{f(a,\tau)}{f(\tau)},
\end{equation}
which according to (\ref{f_marginal}) and (\ref{f_scaling}) behaves as
\begin{equation}
f(a\vert \tau)\approx\frac{1}{2^{1/2}\tau^{3/2}}f_{\textrm{Airy}}\left(\frac{a}{2^{1/2}\tau^{3/2}}\right),\qquad \tau\rightarrow\infty.
\end{equation}
Taking into account that the variance $\sigma^2$ of the double exponential jump distribution equals $2$, we can write the above expression as
\begin{equation}
f(a\vert \tau)\approx\frac{1}{\sigma \tau^{3/2}}f_{\textrm{Airy}}\left(\frac{a}{\sigma \tau^{3/2}}\right),\quad \tau\rightarrow\infty.
\end{equation}
This limit was first proved by Tak\'{a}cs for simple lattice random walks \cite{Takacs91} and was recently extended to all symmetric lattice random walks with a finite variance $\sigma$ \cite{Denisov15}. Here we showed that the scaling limit also holds for the random walk on a real line with the double exponential jump distribution.

\appendix
\setcounter{section}{2}
\section{Analytic properties of \texorpdfstring{$\mathbb{E}_{\omega}[\tau](p,z)$}{E[tau](p,z)}  and \texorpdfstring{$\mathbb{E}_{\omega}[a](p,z)$}{E[a](p,z)}}
\label{appendix_c}

\subsection{Proof that \texorpdfstring{$\mathbb{E}_{\omega}[\tau](p,z)=\mu$}{E[tau](p,z)=mu} has a solution for any \texorpdfstring{$\mu\geq 1$}{mu>1}}

Let us recall the definition of the function $\mathbb{E}[\tau]_{\omega}(p,z)$
\begin{equation}
\mathbb{E}[\tau]_{\omega}(p,z)=\frac{\int_{0}^{\infty}\rmd a \sum_{\tau=1}^{\infty}\tau f(a,\tau)z^{\tau}\rme^{-pa}}{g(p,z)}
\end{equation}
The function $\mathbb{E}_{\omega}[\tau](p,z)$ is monotonically increasing in $z$ for a fixed $p$, which can be seen by inspecting its first derivative which is always non-negative 
\begin{equation}
\frac{\partial}{\partial z}\mathbb{E}_{\omega}[\tau]=\frac{\mathbb{E}_{\omega}[\tau^2]-\mathbb{E}_{\omega}[\tau]^2}{z}\geq 0.
\end{equation}

\begin{figure}[hbt]
	\centering\includegraphics[width=10cm]{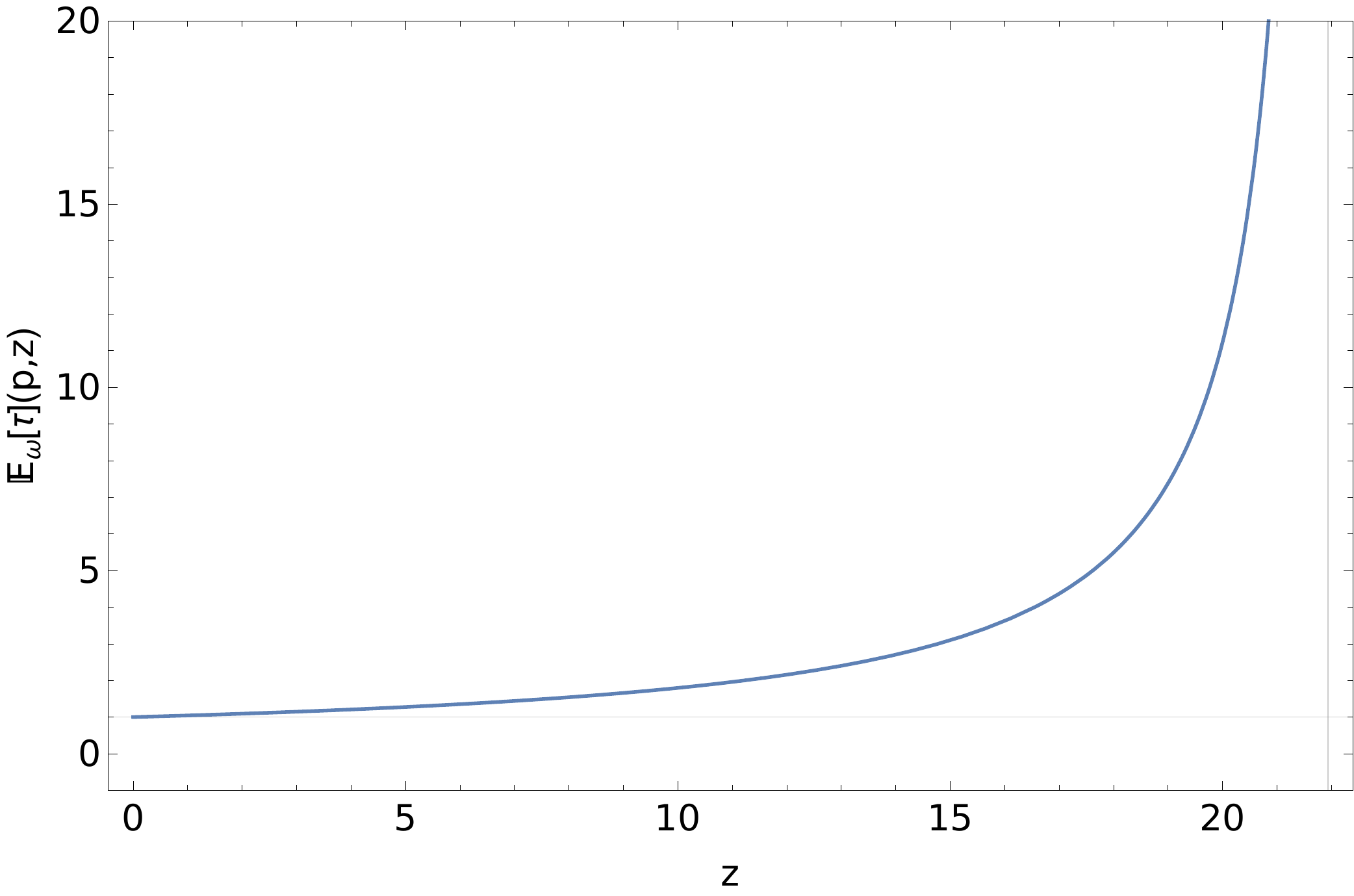}
	\caption{$\mathbb{E}_{\omega}[\tau](p,z)$ plotted as a function of $z$ for fixed $p=10$. The function is monotonically increasing from $1$ to $\infty$ for any $p\geq 0$.}
	\label{figB1}
\end{figure}

To inspect the limits of $\mathbb{E}_f[\omega](z,p)$, we write the expression for $\mathbb{E}_{\omega}[\tau](p,z)$ using (\ref{transform_2}) which yields
\begin{equation}
\fl\qquad\mathbb{E}[\tau](z,p)=1-\frac{2}{p}+\frac{z^{1/2}}{p}\left[\frac{J_{2/p-1}(2z^{1/2}/p)}{J_{2/p}(2z^{1/2}/p)}+\frac{J_{2/p}(2z^{1/2}/p)}{J_{2/p-1}(2z^{1/2}/p)}\right].
\end{equation}
Clearly the function $\mathbb{E}_{\omega}[\tau](p,z)$ is analytic for $0\leq z<(pj_{2/p-1,1}/2)^2$ and diverges to $\infty$ for $z=(pj_{2/p-1,1}/2)^2$. The expansion around $z=0$ reads,
\begin{equation}
\mathbb{E}_{\omega}[\tau](p,z)=1+\frac{1}{2(p+2)}z+\Or(z^2),
\end{equation}
so that $\mathbb{E}[\tau](p,z=0)=1$. This completes our proof that the equation $\mathbb{E}_{\omega}[\tau](p,z)=\mu$ has a unique solution for any $\mu\geq 1$. As an example, we plotted the function $\mathbb{E}_{\omega}[\tau](p,z)$ for fixed $p=10$ in Figure \ref{figB1}.

\subsection{Analytic properties of \texorpdfstring{$\mathbb{E}_{\omega}[a](p,z_0(p,\mu))$}{E[a](p,z0(p,mu))}}

Let us recall the definition of the function $\mathbb{E}[a]_{\omega}(p,z)$
\begin{equation}
\mathbb{E}_{\omega}[a](p,z)=\frac{\int_{0}^{\infty}\rmd a \sum_{\tau=1}^{\infty}a f(a,\tau)z^{\tau}\rme^{-pa}}{g(p,z)}.
\end{equation}
As a function of $p$, $\mathbb{E}_{\omega}[a](p,z_0(p;\mu))$ is monotonically decreasing since
\begin{equation}
\fl\quad \frac{\rmd}{\rmd p}\mathbb{E}_{\omega}[a](p,z_0(p;\mu))=\frac{\mathbb{E}_{\omega}[(a-\sigma)(\tau-\mu)]^2-\mathbb{E}_{\omega}[(a-\sigma)^2]\mathbb{E}_{\omega}[(\tau-\mu)^2]}{\mathbb{E}_{\omega}[(\tau-\mu)^2]}\leq 0,
\end{equation}
which is by the Cauchy-Schwartz inequality always negative. 

To inspect the limit $p\rightarrow\infty$ of $\mathbb{E}_{\omega}[a](p,z_0(p,\mu))$, we first obtain the expression for $g(p,z)$ for fixed $z$ and large $p$,
\begin{equation}
g(p,z)\approx \frac{zp}{2p-z},\qquad p\rightarrow\infty
\label{g-large-p}
\end{equation}
which one gets by expanding Bessel functions $J_{2/p}(2z^{1/2}/p)$ and $J_{2/p-1}(2z^{1/2}/p)$ for small argument. The solution $z_0(p,\mu)$ to the equation $\mathbb{E}_{\omega}[\tau](p,z_0)=\mu$ is then given by
\begin{equation}
z_0(p,\mu)\approx 2\left(1-\frac{1}{\mu}\right)p,\qquad p\rightarrow\infty.
\label{z0-large-p}
\end{equation}
Inserting $z_0(p,\mu)$ into $g(p,z_0)$ and then in the expression for the function $\mathbb{E}_{\omega}[a](p,z)$ gives finally
\begin{equation}
\mathbb{E}_{\omega}[a](p,z_0(p,\mu))\approx \frac{1}{p}, \qquad p\rightarrow\infty,
\end{equation}
which goes to zero when $p$ goes to infinity.

%------------------------------------------------
% REFERENCES
%------------------------------------------------


\vspace*{2ex}
\begin{thebibliography}{99}

\bibitem{Ellis95} Ellis R S 1995 \textit{Scand. Actuarial J.} {\bf 1} 97--142

\bibitem{Touchette09} Touchette H 2009 \textit{Phys. Rep.} \textbf{478} 1--69, arXiv:0804.0327 [cond-mat.stat-mech]

\bibitem{BDGJL05} Bertini L, De Sole A, Gabrielli D, Jona-Lasinio G and Landim C 2005 {\it Phys. Rev. Lett.} {\bf 94} 030601, arXiv:cond-mat/0407161 

\bibitem{BD05} Bodineau T and Derrida B 2005 \textit{Phys. Rev. E} \textbf{72} 066110, arXiv:cond-mat/0506540

\bibitem{Derrida07} Derrida B 2007 {\it J. Stat. Mech.} P07023, arXiv:cond-mat/0703762

\bibitem{EH05} Evans M R and Hanney T 2005 {\it J. Phys. A: Math. Gen.} {\bf 38} R195, arXiv:cond-mat/0501338

\bibitem{Zannetti14} Zannetti M, Corberi F and Gonnella G 2014 \textit{Phys. Rev. E} {\bf 90} 012143

\bibitem{Zannetti15}  Zannetti M 2015 \textit{EPL} {\bf 111}(2)

\bibitem{Linnik61} Linnik Yu V 1961 {\it Proceedings of the 4th Berkeley Symposium on Mathematical Statistics and Probability} vol 2 (London: Cambridge University Press) p~289

\bibitem{Nagaev69} Nagaev A V 1969 {\it Theory Probab. Appl.} {\bf 14} 51

\bibitem{regular-variation} Bingham N H, Goldie C M and Teugels J L 1987 {\it Regular Variation} (Cambridge: Cambridge University Press) p 334

\bibitem{UchaikinZolotarev} Uchaikin V V and Zolotarev V M 1999 {\it Chance and Stability: Stable Distributions and their Applications} (Utrecht: VSP) p~144

\bibitem{Tribelsky02} Tribelsky M I (2002) {\it Phys. Rev. Lett.} {\bf 89} 070201

\bibitem{MEZ05} Majumdar S N, Evans M R and Zia R K P 2005 {\it Phys. Rev. Lett.} {\bf 94} 180601, arXiv:cond-mat/0501055

\bibitem{EMZ06} Evans M R, Majumdar S N and Zia R K P 2006 {\it J. Stat. Phys.} {\bf 123} 357--90, arXiv:cond-mat/0510512 [cond-mat.stat-mech]

\bibitem{MKB98} Majumdar S N, Krishnamurthy S, Barma M 1998 {\it Phys. Rev. Lett.} {\bf 81} 3691, arXiv:cond-mat/9806353

\bibitem{GSS03}  Gro{\ss}kinsky S, Sch\"{u}tz G M and Spohn H 2003 {\it J. Stat. Phys.} {\bf 113} 389--410, arXiv:cond-mat/0302079

\bibitem{HMS09} Hirschberg O, Mukamel D and Sch\"utz G M 2009 {\it Phys. Rev. Lett.} {\bf 103} 090602, arXiv:0906.0709 [cond-mat.stat-mech]

\bibitem{AL11} Armend\'{a}riz I and Loulakis M 2011 {\it Stoch. Proc. Appl.} {\bf 121}(5) 1138--1147, arXiv:0912.1516 [math.PR]

\bibitem{AGL13} Armend\'{a}riz I, Grosskinsky S  and Loulakis M 2013 {\it Stoch. Proc. Appl.} {\bf 123} 3466, arXiv:0912.1793 [math.PR]

\bibitem{WCEB14} Whitehouse J, Costa A, Blythe R A and Evans M R 2014 {\it J. Stat. Mech.}  P11029, arXiv:1301.2489 [cond-mat.stat-mech]

\bibitem{EW14} Evans M R and Waclaw B 2014 {\it J. Phys. A: Math. Theor.} {\bf 47} 095001, arXiv:1312.5642 [cond-mat.stat-mech]

\bibitem{SNEM14a} Szavits-Nossan J, Evans M R and Majumdar S N 2014 {\it Phys. Rev. Lett.} {\bf 112} 020602, arXiv:1309.4255 [cond-mat.stat-mech]

\bibitem{SNEM14b} Szavits-Nossan J, Evans M R and Majumdar S N 2014 {\it J. Phys. A: Math. Theor.} {\bf 47} 455004, arXiv:1406.3573 [cond-mat.stat-mech]

\bibitem{SM06} Schehr G and Majumdar, S N 2006 {\it Phys. Rev. E} {\bf 73} 056103, arXiv:cond-mat/0601073

\bibitem{EHM06}  Evans M R, Hanney T and Majumdar S N 2006 {\it Phys. Rev. Lett.} {\bf 97} 010602, arXiv:cond-mat/0604664

\bibitem{CT13} Chetrite R and Touchette H 2013 {\it Phys. Rev. Lett.} {\bf 111} 120601, arXiv:1306.4563 [cond-mat.stat-mech]

\bibitem{CT15} Chetrite R and Touchette H  2015 {\it Ann. Henri Poincar\'e} {\bf 16} 2005--2057, arXiv:1405.5157 [cond-mat.stat-mech]

\bibitem{Frey15} Knebel J, Weber M F, Kr\"{u}ger T and Frey E 2015 {\it Nat. Commun.} {\bf 6} 6977

\bibitem{SNE15} Szavits-Nossan J and Evans M R  2015 {\it J. Stat. Mech.} P12008, arXiv:1508.04969 [cond-mat.stat-mech]

\bibitem{Lindley52} Lindley D V 1952 {\it Mathematical Proceedings of the Cambridge Philosophical Society} {\bf 48} 277--289

\bibitem{Asmussen03} Asmussen S 2003 {\it Applied Probability and Queues} (New York: Springer-Verlag) p~92

\bibitem{Takacs91} Tak\'acs L 1991 {\it Adv. Appl. Prob.} {\bf 23} 557--585

\bibitem{KM05} Kearney M J and Majumdar S N 2005 {\it J. Phys. A: Math. Gen.} {\bf 38} 4097, arXiv:cond-mat/0501445

\bibitem{M05} Majumdar S N 2005 {\it Current Science} {\bf 89} 2076, arXiv:cond-mat/0510064

\bibitem{M10} Majumdar S N 2010 {\it Physica A} {\bf 389} 4299, arXiv:0912.2586 [cond-mat.stat-mech]      

\bibitem{SA54} Sparre Andersen E 1954 {\it Math. Scand.} {\bf 2} 195

\bibitem{Feller2} Feller W 1971 {\it An Introduction to Probability Theory and Its Applications II} (New York: Wiley)

\bibitem{BMS13} Bray A J, Majumdar S N and Schehr G 2013 {\it Advances in Physics} {\bf  62} 225, arXiv:1304.1195 [cond-mat.stat-mech]

\bibitem{Olver54} Olver F W J 1954 {\it Phil. Trans. R. Soc. A} {\bf 247}(930) 328--368

\bibitem{Ellis00} Ellis R S, Haven K and Turkington B 2000 {\it J. Stat. Phys.} {\bf 101} 999, arXiv:math/0012081

\bibitem{WSJM09a} Waclaw B, Sopik J, Janke W and Meyer-Ortmanns H 2009 {\it Phys. Rev. Lett.} {\bf 103} 080602, arXiv:0901.3664 [cond-mat.stat-mech]

\bibitem{WSJM09b} Waclaw B, Sopik J, Janke W and Meyer-Ortmanns H 2009 {\it J. Stat. Mech.} {\bf P10021} 

\bibitem{ENJ14}  Ehrenpreis E, Nagel H and Janke W  2014 {\it J. Phys. A: Math. Theor.} {\bf 47} 125001, arXiv:1506.01641 [cond-mat.stat-mech]

\bibitem{Kishore63} Kishore N 1963 {\it Proceedings of the American Mathematical Society} {\bf 14} 527

\bibitem{Kishore64} Kishore N 1964 {\it Duke Math. J.} {\bf 31}(3) 513--518

\bibitem{Olver51} Olver F W J 1951 {\it Proc. Cambridge Philos. Soc.} {\bf 47} 699--712

\bibitem{Denisov15} Denisov D, Kolb M and Wachtel W 2015 {\it J. London Math. Soc.} {\bf 91}(2) 495--513

\bibitem{MC05} Majumdar S N and Comtet A 2005 {\it J. Stat. Phys.} {\bf 119} 777, arXiv:cond-mat/0409566

\bibitem{Kishore64_2} Kishore N 1964 {\it Proceedings of the American Mathematical Society} {\bf 15}(6) 911--917
\end{thebibliography}
\end{document}